\documentclass[twocolumn]{aastex7}
\usepackage{amsmath}
\usepackage{lineno}
\usepackage{comment}
\linenumbers

\begin{document}

\title{Habitable Zones Around Massive Stars: From the Main Sequence to Supergiants}

\author[orcid=0000-0003-1927-4397,sname='Devesh Nandal']{Devesh Nandal}

\affiliation{Center for Astrophysics, Harvard and Smithsonian, 60 Garden St, Cambridge, MA 02138, USA}
\email[show]{deveshnandal@yahoo.com}  

\author[orcid=0000-0003-4330-287X,gname=Bosque, sname='Sur America']{Abraham Loeb} 

\affiliation{Center for Astrophysics, Harvard and Smithsonian, 60 Garden St, Cambridge, MA 02138, USA}
\email{fakeemail2@google.com}

\begin{abstract}
Massive stars dominate the feedback of young stellar populations, yet their ultraviolet fields and winds are often presumed to preclude Earth like habitability. We test this by mapping time dependent habitable zones (HZs) for solar metallicity stars of $0.8$--$120\,M_\odot$. Using rotating and non rotating \textsc{GENEC} tracks, we compute bolometric climate HZ boundaries and impose XUV energy limited escape and wind ram pressure constraints for a dipole-magnetized Earth analogue. The retention limited inner edge is the most restrictive limit. We measure annulus lifetime, longest fixed orbit residence, and maximum dynamically packed terrestrial multiplicity, finding a sharp main-sequence ceiling. A rotating $9\,M_\odot$ star sustains a retention limited HZ for $\sim 30$ Myr at $\sim 70$--$130$ AU, but becomes brief and narrow by $12\,M_\odot$ and disappears by $15\,M_\odot$. Post main-sequence evolution can reopen HZs up to $\sim 25$--$30\,M_\odot$, but only for $\sim 0.03$--$1.5$ Myr at hundreds to $\sim 10^3$ AU, disappearing by $\sim 40\,M_\odot$. Stellar rotation modestly increases habitable lifetimes near the upper main sequence without altering the high mass ceiling. IMF weighting shows that massive stars contribute only $\sim 10^{-4}$ of the habitable planet time budget. Even so, for the fiducial occurrence normalization and if rocky planets form or survive at the required wide separations, they add a few $10^5$ Earth analogues satisfying the adopted criteria to the Milky Way at any instant. Massive star systems do not dominate the Galaxy-wide habitability budget, but may provide short-lived, distinct targets for biosignature searches.

\end{abstract}


\section{Introduction}

Massive stars ($M_\star \gtrsim 8\,M_\odot$) are intrinsically rare in standard initial mass functions \citep[e.g.,][]{Salpeter1955,Kroupa2001,Chabrier2003}, yet they dominate both the luminosity and mechanical power of young stellar populations.
Their far-ultraviolet and ionizing photons shape the radiation environment of star forming regions, while their line driven winds and eventual core-collapse supernovae inject momentum and newly synthesized elements into the interstellar medium \citep[e.g.,][]{Maeder1997,ZinneckerYorke2007,Crowther2007,Krumholz2014,Langer2012}.
Because of these outsized impacts, the structure and evolution of massive stars, including mass loss, rotation, and internal mixing have been studied in detail for decades \citep[e.g.,][]{Castor1975,Kudritzki2002,Vink2001,Meynet2006,Brott2011,Ekstrom2012}.
Modern stellar evolution calculations that incorporate these processes provide time-dependent luminosities, radii, effective temperatures, and mass-loss rates that underpin a wide range of astrophysical applications \citep[e.g.,][]{Eggenberger2008,Ekstrom2012, Nandal2024b}.
Yet one question that has received comparatively little quantitative attention is whether, and for how long, massive stars can host circumstellar environments compatible with surface liquid water and atmospheric retention on terrestrial planets.

The circumstellar habitable zone (HZ) has long served as a pragmatic first filter in the search for potentially habitable worlds \citep[e.g.,][]{Kasting1993,Kopparapu2013,Kopparapu2014,Kaltenegger2021}.
Early discussions of life supporting orbital regions go back to \citet{Huang1959,Huang1960}, with the first systematic numerical HZ boundaries developed by \citet{Hart1979} and the widely adopted modern baseline established by \citet{Kasting1993}.
In its classical climate definition, the HZ is the range of orbital distances where an Earth-like planet with a suitable atmosphere can sustain liquid water on its surface. Over the past decade, both 1D and 3D climate calculations have refined these limits and provided widely used flux-based prescriptions \citep[e.g.,][]{Selsis2007,KalteneggerSasselov2011,Kopparapu2013,Kopparapu2014,Yang2014} 

Because stellar luminosities evolve, HZ boundaries are time dependent; this motivates metrics based on the duration of habitability, including the concept of a continuously habitable zone \citep[e.g.,][]{Rushby2013,DanchiLopez2013}.

Most HZ studies and most observational searches for HZ planets have focused on low- and intermediate-mass stars (FGKM) \citep[e.g.,][]{Gilbert2023}.
This emphasis is natural: such stars are far more numerous \citep[e.g.,][]{Salpeter1955,Kroupa2001,Chabrier2003}, they remain on the main sequence for Gyr, and their AU-scale HZs produce orbital periods accessible to transit and radial-velocity surveys \citep[e.g.,][]{Borucki2010,Rauer2014,Ricker2015}.
By contrast, early-type and massive stars are challenging targets for traditional planet searches because they are often rapid rotators with sparse spectral lines, and because their HZ radii scale outward roughly as $r \propto L_\star^{1/2}$, pushing temperate climates to tens or hundreds of AU \citep{Kasting1993, Kaltenegger2021}.
The combination of wide orbits, strong ultraviolet radiation fields, and short lifetimes has led to the common assumption that massive stars are irrelevant to planetary habitability.

Habitability, however, is not determined by bolometric irradiation alone.
High energy photons and particle outflows can drive rapid atmospheric erosion, modifying (or eliminating) surface habitability even when a planet receives temperate bolometric flux, and such loss processes can leave observable demographic imprints in exoplanet populations \citep[e.g.,][]{Airapetian2020,David2021}.
A useful baseline for XUV-driven hydrodynamic escape is the energy-limited formalism originally developed for Solar System atmospheres \citep{Watson1981} and widely applied to exoplanets \citep[e.g.,][]{Lammer2003,Lecavelier2007,OwenJackson2012, Forbes2018}.
Stellar winds can also strip atmospheres and compress magnetospheres; in that context, magnetic shielding and pressure balance arguments date back to early solar-wind/magnetosphere theory \citep{ChapmanFerraro1931} and have been extended to exoplanetary environments \citep[e.g.,][]{Griessmeier2004,Vidotto2013, Pezzotti2025}.
These non-climate constraints become especially acute for massive stars, whose radiative output is concentrated at short wavelengths and whose line driven winds can exceed solar values by orders of magnitude \citep[e.g.,][]{Parker1958,Castor1975,Vink2001,Puls2008}.

Beyond atmospheric loss, the formation and survival of planets on the wide orbits implied by massive star HZs is itself uncertain \citep[e.g.,][]{Winter2022}.
Disk masses generally increase with stellar mass, providing more raw material for planet formation \citep[e.g.,][]{Andrews2013,Williams2011}, but external irradiation from nearby OB stars can photoevaporate and truncate disks in clustered environments \citep[e.g.,][]{Johnstone1998}.
At the tens to hundreds of AU separations relevant here, rocky planet formation and long term survival should therefore be regarded as conditional assumptions rather than outcomes predicted by this work. Existing work that incorporates stellar evolution has largely targeted Sun-like and lower-mass stars, or the post-main-sequence expansion of HZs around subgiants and giants \citep[e.g.,][]{Lopez2005,DanchiLopez2013,Ramirez2016}.
To our knowledge, there has not been a systematic mapping of whether massive stars can host a retention-limited habitable annulus once atmospheric-retention constraints from XUV irradiation and winds are coupled to realistic massive-star evolutionary tracks across both main-sequence and post-main-sequence phases.

In this paper, we connect massive star evolution to terrestrial planet habitability by computing time-dependent HZs for stars spanning $0.7$--$120\,M_\odot$ at solar metallicity.
We use Geneva \textsc{GENEC} evolutionary tracks with and without rotation \citep{Eggenberger2008,Ekstrom2012} and define a retention-limited HZ whose outer edge follows standard climate limits while whose inner edge is set by the most restrictive of three criteria: (i) the bolometric inner climate limit, (ii) XUV-driven energy-limited atmospheric escape, and (iii) wind ram-pressure erosion moderated by a dipolar magnetosphere.
We quantify the cumulative time a retention-limited annulus exists, the maximum continuous residence time for a planet on a fixed orbit, and characteristic radii at which habitability is most likely during each evolutionary phase.
Motivated by the large radial extent of HZs around luminous stars, we also estimate the maximum dynamical multiplicity of terrestrial planets that can be packed into the annulus and fold these results through Milky-Way-like initial mass functions to evaluate the population-weighted contribution of massive stars.
Our results show that main-sequence habitability under the retention-limited criteria terminates near $M_\star \sim 10$--$15\,M_\odot$ depending on the residence criterion, with post-main-sequence evolution briefly reopening habitable annuli at higher masses, and that the mass weighted contribution of massive stars is negligible even though the absolute number of such systems in a Milky Way like galaxy can be large.

This paper is organized as follows.
In Section~\ref{sec:methods} we describe the stellar evolution tracks, the climate and atmospheric-retention criteria, and our diagnostics for habitability and packed multiplicity.
Section~\ref{sec:results} presents the resulting time-dependent HZ maps, residence-time and multiplicity estimates, and IMF-weighted yields.
In Section~\ref{sec:discussion} we discuss physical interpretations, limitations, and observational implications.
Section~\ref{sec:conclusion} summarizes our main conclusions.

\section{Methods}\label{sec:methods}

\subsection{Stellar evolution tracks}\label{subsec:tracks}
Stellar evolution tracks at solar metallicity $Z=0.014$ are computed with the Geneva Stellar Evolution Code (\textsc{GENEC}; \citep{Eggenberger2008,Nandal2023, Nandal2024b, Nandal2025b}. The adopted initial rotation is $v_{\rm ini}/v_{\rm crit}=0.4$, consistent with the standard Geneva rotating solar-metallicity grids (\citealt{Ekstrom2012}). The model set spans $M_{\rm ini}=0.8$--$120\,M_\odot$. Tracks are analysed from the ZAMS to core helium exhaustion. If a track extends beyond that point then it is truncated at core He exhaustion for uniformity. Later phases are shorter, more model dependent, and more strongly affected by late mass loss and endpoint physics, so they are not used in the uniform comparison. The end time is denoted $t_{\rm end}$. The final mass is $M_{\rm end}\equiv M_\star(t_{\rm end})$. The integrated mass loss is $\Delta M\equiv M_{\rm ini}-M_{\rm end}$. Values of $(t_{\rm end},M_{\rm end},\Delta M)$ are taken directly from the track output and summarised in Table~\ref{tab:1}. All tracks are single star models; binary mass transfer, mergers, common envelope evolution, and binary modified wind or irradiation histories are not included.

The habitable zone (HZ) model uses the time series of stellar age $t$, bolometric luminosity $L_\star(t)$, effective temperature $T_{\rm eff}(t)$, stellar mass $M_\star(t)$, central hydrogen mass fraction $X_{c,{\rm H}}(t)$, and mass-loss rate $\dot{M}(t)$. The photospheric radius is inferred from
\begin{equation}
R_\star(t)=\left[\frac{L_\star(t)}{4\pi\sigma_{\rm SB}T_{\rm eff}(t)^4}\right]^{1/2},
\label{eq:Rstar}
\end{equation}
where $\sigma_{\rm SB}$ is the Stefan--Boltzmann constant. 

We parameterise the wind terminal speed as a temperature dependent multiple of the escape speed,
\begin{equation}
v_{\infty}(t)=\eta_{\infty}\!\left[T_{\rm eff}(t)\right]\,v_{\rm esc}(t).
\label{eq:vinf}
\end{equation}
The escape speed is
\begin{equation}
v_{\rm esc}(t)=\left(\frac{2GM_{\star}(t)}{R_{\star}(t)}\right)^{1/2}.
\label{eq:vesc}
\end{equation}
To mimic the usual bi-stability behaviour we adopt a two-branch scaling \citep{Vink2001},
\begin{equation}
\eta_{\infty}(T_{\rm eff})=
\begin{cases}
\eta_{\rm hot}, & T_{\rm eff}\ge T_{\rm bist},\\
\eta_{\rm cool}, & T_{\rm eff}< T_{\rm bist}.
\end{cases}
\label{eq:eta_inf}
\end{equation}
We take $\eta_{\rm hot}=2.6$, $\eta_{\rm cool}=1.3$, and $T_{\rm bist}=2.1\times10^{4}\,{\rm K}$. Main-sequence (MS) timesteps satisfy $X_{c,{\rm H}}\ge X_{\rm crit}$ with $X_{\rm crit}=0.01$. Post-MS timesteps satisfy $X_{c,{\rm H}}<X_{\rm crit}$. This split is used when reporting HZ windows and multiplicities.

\subsection{Habitable-zone model}\label{subsec:hzmodel}
At each timestep the HZ is defined as an annulus $[r_{\rm in}(t),r_{\rm out}(t)]$ in astronomical units (AU). The outer edge is set by a bolometric ``climate'' criterion. The inner edge is the maximum of three constraints. These constraints are a climate inner edge, an XUV-driven atmospheric-loss edge, and a wind-pressure edge.

\subsubsection{Climate HZ}\label{subsubsec:climate}
The climate HZ is implemented as a pure bolometric flux scaling with fixed effective flux thresholds \citep{Kopparapu2013}. The inner and outer climate edges are
\begin{equation}
\begin{aligned}
r_{\rm in,clim}(t)  &= \left[\frac{L_\star(t)/L_\odot}{S_{\rm eff,in}}\right]^{1/2},\\
r_{\rm out,clim}(t) &= \left[\frac{L_\star(t)/L_\odot}{S_{\rm eff,out}}\right]^{1/2},
\end{aligned}
\label{eq:climate}
\end{equation}
where $L_\odot$ is the solar luminosity. The adopted constants are $S_{\rm eff,in}=1.015$ and $S_{\rm eff,out}=0.35$. These are treated as fixed surrogates. No explicit spectral correction with $T_{\rm eff}$ is applied. The adopted outer edge is
\begin{equation}
r_{\rm out}(t)=r_{\rm out,clim}(t).
\label{eq:rout}
\end{equation}

We use fixed bolometric thresholds as the fiducial climate prescription because the temperature-dependent polynomial corrections of \citet{Kopparapu2013,Kopparapu2014} are calibrated for cooler stars and do not extend to the OB-star effective temperatures reached by much of our grid. We therefore treat these thresholds as transparent bolometric surrogates and test the sensitivity to this choice in Section~\ref{subsec:climate_sensitivity}.

\subsubsection{XUV atmospheric-loss inner edge}\label{subsubsec:xuv}
The XUV constraint is evaluated for an Earth analog planet with mass $M_p=M_\oplus$ and radius $R_p=R_\oplus$. The atmosphere mass is fixed to $M_{\rm atm}=5\times10^{18}\,\mathrm{kg}$. The heating efficiency is $\varepsilon$. The exposure time is $\tau$. A tidal correction factor is set to $K_{\rm tide}=1$.

A blackbody approximation is used to estimate the stellar XUV luminosity. The XUV fraction is
\begin{equation}
f_{\rm XUV}(T_{\rm eff})=
\frac{\int_{\lambda_{\min}}^{\lambda_{\max}} B_\lambda(\lambda,T_{\rm eff})\,\mathrm{d}\lambda}
{\sigma_{\rm SB}T_{\rm eff}^4/\pi},
\label{eq:fxuv}
\end{equation}
where $B_\lambda$ is the Planck function. The bandpass is $\lambda_{\min}=10\,\mathrm{nm}$ to $\lambda_{\max}=118\,\mathrm{nm}$. The upper edge is chosen to include the broad EUV and XUV heating band while remaining below Ly$\alpha$, rather than to isolate only photons shortward of the hydrogen ionization edge at $91.2\,\mathrm{nm}$. The sensitivity of the result to the effective high-energy normalization is tested in Section~\ref{subsec:xuv_sensitivity}. The corresponding XUV luminosity is
\begin{equation}
L_{\rm XUV}(t)=4\pi R_\star(t)^2\,\sigma_{\rm SB}T_{\rm eff}(t)^4\,f_{\rm XUV}[T_{\rm eff}(t)].
\label{eq:Lxuv}
\end{equation}
The XUV inner edge is defined from an integrated energy-limited loss threshold \citep{Watson1981},
\begin{equation}
\begin{split}
r_{\rm XUV}(t)=\frac{1}{\mathrm{AU}}
\left[\frac{\varepsilon\,R_p^3\,L_{\rm XUV}(t)\,\tau}
{4\,G\,M_p\,K_{\rm tide}\,M_{\rm atm}}\right]^{1/2}.
\end{split}
\label{eq:rXUV}
\end{equation}
where $\varepsilon$ is the heating efficiency, $R_p$ and $M_p$ are the planet radius and mass, $L_{\rm XUV}(t)$ is the stellar XUV luminosity, $\tau$ is the exposure time, $G$ is the gravitational constant, $K_{\rm tide}$ is a tidal correction factor, $M_{\rm atm}$ is the atmospheric mass to be removed, and $\mathrm{AU}$ is the astronomical unit.

We use this Planck-function estimate of $L_{\rm XUV}$ as a fiducial normalization rather than as a calibrated stellar XUV spectrum. Equation~\ref{eq:rXUV} should be interpreted as a retention screening criterion for the removal of one Earth atmosphere over the adopted exposure time $\tau$, not as a self-consistent atmospheric evolution model. Because $r_{\rm XUV}$ depends on the product $L_{\rm XUV}\tau$, the sensitivity test in Section~\ref{subsec:xuv_sensitivity} also brackets the adopted exposure time.

In a self-consistent atmospheric evolution calculation, the instantaneous escape rate would be evaluated along the stellar track and integrated at fixed orbit,
\begin{equation}
\Delta M_{\rm esc}(a)=
\int_{t_1}^{t_2}\dot{M}_{\rm esc}(t,a)\,\mathrm{d}t .
\label{eq:integrated_escape}
\end{equation}
This cumulative loss would then be compared with the atmospheric inventory as the climate and retention boundaries evolve. Such a treatment would couple atmospheric escape directly to the residence-time calculation, because the relevant exposure time would be the actual time spent by the planet inside the evolving HZ. The present work does not attempt this coupled evolution. Instead, Equation~\ref{eq:rXUV} defines a transparent screening boundary, and Section~\ref{subsec:hz_residence} separately measures the longest fixed-orbit residence time inside the resulting retention-limited annulus.

\subsubsection{Wind pressure inner edge with dipole scaling}\label{subsubsec:wind}
A wind truncation is computed from ram pressure balance with a dipolar planetary magnetic field. The equatorial surface field is $B_p$. The critical magnetopause distance is $R_{\rm mp}=R_{\rm crit}R_p$ with dimensionless $R_{\rm crit}$. Dipole scaling gives the field at the magnetopause,
\begin{equation}
B_{\rm mp}=\frac{B_p}{R_{\rm crit}^3}.
\label{eq:Bmp}
\end{equation}
The maximum magnetic pressure is
\begin{equation}
P_{\max}=\frac{B_{\rm mp}^2}{2\mu_0}=\frac{B_p^2}{2\mu_0\,R_{\rm crit}^6},
\label{eq:Pmax}
\end{equation}
with vacuum permeability $\mu_0$. The wind ram pressure at orbital distance $a$ is approximated as
\begin{equation}
P_{\rm w}(a,t)=\frac{\dot{M}(t)\,v_\infty(t)}{4\pi a^2},
\label{eq:Pwind}
\end{equation}
where $\dot{M}$ is converted to SI units and $v_\infty$ follows Equation~\ref{eq:vinf}. The wind inner edge follows from $P_{\rm w}=P_{\max}$,
\begin{equation}
r_{\rm wind}(t)=\frac{1}{\mathrm{AU}}
\left[\frac{\dot{M}(t)\,v_\infty(t)}{4\pi P_{\max}}\right]^{1/2}.
\label{eq:rwind}
\end{equation}
The fiducial values are $B_p=0.3\,\mathrm{G}$ and $R_{\rm crit}=2.5$. The wind term uses the smooth-wind ram pressure and an Earth-analogue dipole field as a retention-screening prescription. From Equations~\ref{eq:Pmax} and \ref{eq:rwind},
\begin{equation}
r_{\rm wind}\propto
\left(\dot{M}v_\infty\right)^{1/2}B_p^{-1}R_{\rm crit}^{3}.
\label{eq:rwind_scaling}
\end{equation}
Thus an enhancement of the instantaneous wind ram pressure by a factor $f_{\rm cl}$ would move the wind boundary outward as $r_{\rm wind}\rightarrow f_{\rm cl}^{1/2}r_{\rm wind}$. The calculation should therefore be interpreted as a mean-wind retention threshold, not as a model of clumped wind structure or planetary dynamo evolution.

\subsubsection{Retention-limited HZ and scenarios}\label{subsubsec:operational}
The inner edge of the retention-limited HZ is
\begin{equation}
r_{\rm in}(t)=\max\!\left[r_{\rm in,clim}(t),\,r_{\rm XUV}(t),\,r_{\rm wind}(t)\right].
\label{eq:rin}
\end{equation}
A habitable annulus exists if $r_{\rm in}(t)<r_{\rm out}(t)$ with both edges finite and positive. Two reduced scenarios are also reported. The climate-only case sets $r_{\rm in}=r_{\rm in,clim}$. The climate+XUV case sets $r_{\rm in}=\max(r_{\rm in,clim},r_{\rm XUV})$.

\subsection{Time-domain diagnostics}\label{subsec:timediag}
A Boolean mask $\mathcal{H}(t)$ is defined for timesteps satisfying $r_{\rm in}(t)<r_{\rm out}(t)$. The longest continuous habitable interval is the maximum contiguous time segment for which $\mathcal{H}(t)$ holds. This is evaluated separately on the MS and post-MS.

A representative epoch is defined to report a single ``widest-HZ'' geometry. Within a given phase, the logarithmic width
\begin{equation}
\Delta\ln r(t)=\ln\!\left[\frac{r_{\rm out}(t)}{r_{\rm in}(t)}\right]
\label{eq:dlnr}
\end{equation}
is maximised over timesteps with $\mathcal{H}(t)=\mathrm{true}$. The maximising time is denoted $t_\star$. When a model has any MS HZ timesteps, $t_\star$ is taken on the MS. Otherwise it is taken on the post-MS. The radii $(r_{\rm in},r_{\rm out})$ evaluated at $t_\star$ define the annulus used for the headline multiplicity estimates.

\subsection{Planet multiplicity in the HZ}\label{subsec:mult}
Two multiplicity estimators map an annulus $[r_{\rm in},r_{\rm out}]$ into an upper bound on the number of planets.

Our multiplicity calculation is intended as an upper bound on the dynamical capacity of a retention-limited annulus, not as a new $N$-body stability study. The problem of packing terrestrial planets inside the HZ has been studied with direct integrations and stability maps \citep[e.g.,][]{Obertas2017,Agnew2019,Kane2020}. We therefore use two simple estimators that map the time-dependent annulus into a characteristic maximum occupancy, while leaving detailed architecture-dependent stability to future work.

\subsubsection{Model A: minimum period-ratio packing}\label{subsubsec:multA}
A geometric packing bound is obtained by enforcing a minimum adjacent period ratio $\mathcal{R}=P_{i+1}/P_i$. Kepler scaling gives a minimum semimajor-axis ratio $\alpha_{\min}=\mathcal{R}^{2/3}$. The maximum number of planets is
\begin{equation}
N_{\max}=
\begin{cases}
0, & r_{\rm out}\le r_{\rm in},\\
1+\left\lfloor\frac{\ln(r_{\rm out}/r_{\rm in})}{\ln \alpha_{\min}}\right\rfloor, & r_{\rm out}>r_{\rm in}.
\end{cases}
\label{eq:Nratio}
\end{equation}
The fiducial choice is $\mathcal{R}=1.33$, motivated by N-body instability experiments indicating that unstable multiplanet systems typically include at least one adjacent pair with $P_{i+1}/P_i<1.33$ \citep{Wu2019}.

\subsubsection{Model B: mutual-Hill spacing with a solids budget}\label{subsubsec:multB}
A second estimator adds a dynamical spacing condition and a disk-mass budget. Equal-mass planets are assumed. Adjacent planets are required to be separated by $K$ mutual Hill radii \citep{Gladman1993, Chambers1996}. This is written as an approximate constant spacing ratio
\begin{equation}
\mu=\left(\frac{2M_{\rm p}}{3M_\star}\right)^{1/3},\qquad
\alpha=\frac{K}{2}\mu,\qquad
\gamma=\frac{1+\alpha}{1-\alpha},
\label{eq:gamma}
\end{equation}
which is valid for $\alpha<1$. The spacing-only bound is
\begin{equation}
N_{\rm space}=1+\left\lfloor\frac{\ln(r_{\rm out}/r_{\rm in})}{\ln\gamma}\right\rfloor.
\label{eq:Nspace}
\end{equation}

The available solids reservoir is parameterised as
\begin{equation}
M_{\rm dust}=M_{\rm dust,\odot}\left(\frac{M_\star}{M_\odot}\right),
\label{eq:Mdust}
\end{equation}
with $M_{\rm dust,\odot}=50\,M_\oplus$. A rocky fraction $f_{\rm rock}=0.5$ is adopted. The fraction of solids inside the HZ is computed from a radial weight $w(a)\propto a^p$ over $[a_{\min},a_{\max}]$,
\begin{equation}
\begin{split}
f_{\rm HZ}=
\frac{\int_{r_{\rm in}}^{r_{\rm out}} a^p\,\mathrm{d}a}
{\int_{a_{\min}}^{a_{\max}} a^p\,\mathrm{d}a},
\end{split}
\label{eq:fHZ}
\end{equation}
with $p=0$, $a_{\min}=0.1\,\mathrm{AU}$, and $a_{\max}=100\,\mathrm{AU}$. This disk prescription is intentionally simple and should be interpreted as a capacity model. Observed disk masses have large scatter, and the flat radial weighting is optimistic for the wide separations relevant to massive star HZs. The mass available to planets is
\begin{equation}
M_{\rm avail}=\varepsilon_{\rm form}\,f_{\rm rock}\,f_{\rm HZ}\,M_{\rm dust},
\label{eq:Mavail}
\end{equation}
with $\varepsilon_{\rm form}=0.5$. A minimum planet mass is imposed as $M_{\min}=0.1\,M_\oplus$. For each trial multiplicity $N$, the planet mass is set to
\begin{equation}
M_{\rm p}(N)=\max\!\left(M_{\min},\frac{M_{\rm avail}}{N}\right).
\label{eq:Mp}
\end{equation}
The reported multiplicity is the largest integer $N$ that satisfies both $N\le \lfloor M_{\rm avail}/M_{\min}\rfloor$ and $N\le N_{\rm space}[M_{\rm p}(N)]$. The explored stability parameters are $K\in\{12,16,20\}$ and the adopted disk parameters are fiducial. Sensitivity is quantified by varying $f_{\rm rock}$ and $\varepsilon_{\rm form}$ by $\pm30\%$ and repeating the solve.

\subsection{Parameter exploration}\label{subsec:grid}
Uncertainties in atmospheric escape and magnetospheric protection are explored on discrete grids. The exposure time takes $\tau/{\rm Myr}\in\{0.01,0.03,0.1,0.3,1,3,10\}$, the heating efficiency takes $\varepsilon\in\{0.05,0.1,0.3\}$, the surface field takes $B_p/{\rm G}\in\{0.1,0.3,1.0\}$, and the critical standoff distance takes $R_{\rm crit}\in\{2.0,2.5,5.0\}$. The variations in $B_p$ and $R_{\rm crit}$ should be read as shielding brackets for the adopted Earth-analogue magnetosphere, not as a model of planetary dynamo evolution. For each track and each parameter tuple, the radii $r_{\rm XUV}(t)$ and $r_{\rm wind}(t)$ are recomputed, then propagated into $r_{\rm in}(t)$ through Equation~\ref{eq:rin}. Summary products include MS and post-MS habitable intervals, representative radii at $t_\star$, and multiplicity--mass relations under Models A and B.

\subsection{IMF-weighted habitability yield}\label{subsec:imf_yield}
To translate single-track habitable-zone (HZ) diagnostics into population-weighted expectations, we define a time-integrated habitability yield for each stellar model and multiplicity prescription.
For a given track we evaluate the instantaneous packing-limited multiplicity $N_m(t)$ using Method~$m\in\{\mathrm{A},\mathrm{B}\}$ applied to the instantaneous annulus $[r_{\rm in}(t),\,r_{\rm out,clim}(t)]$, and we set $N_m(t)=0$ whenever no annulus exists ($r_{\rm in}\ge r_{\rm out,clim}$).
For Method~B, $N_{\rm B}(t)$ additionally depends on the adopted stability and disk-scale parameters (e.g., $K$ and $a_{\max}$).

The yield is then
\begin{equation}
Y_m(M_{\rm ini}) = \int N_m(t)\,H(t)\,{\rm d}t,
\label{eq:Ym}
\end{equation}
which has units of planet--time (we report planet--Myr) and is evaluated over MS and post-MS phases consistently with the masks defined in Sections~\ref{subsec:timediag} and \ref{subsec:mult}.

We then fold these per-mass yields through an initial mass function (IMF) $\xi(M)$ and report the yield per unit stellar mass formed,
\begin{equation}
\overline{Y}_m = \frac{\int Y_m(M)\,\xi(M)\,{\rm d}M}{\int M\,\xi(M)\,{\rm d}M},
\label{eq:Ybarm}
\end{equation}
together with the cumulative contribution from stars above a threshold mass $M_{\rm cut}$,
\begin{equation}
f_{\ge M_{\rm cut},m}=
\frac{\int_{M_{\rm cut}} Y_m(M)\,\xi(M)\,{\rm d}M}{\int Y_m(M)\,\xi(M)\,{\rm d}M}.
\label{eq:f_ge_Mcut_m}
\end{equation}
Throughout, we compute these quantities over the mass interval common to both grids so that rotating (S0.4) and non-rotating (S0) results can be compared without extrapolation.

\subsection{Milky Way normalization of the IMF-integrated yield}
\label{sec:mw_norm}
Our IMF-integrated yields are reported as \(\overline{Y}_m(M_{\rm min})\), the habitable planet--time produced per unit stellar mass formed, with units of planet--Myr~\(M_\odot^{-1}\), for multiplicity Method \(m\in\{\mathrm{A},\mathrm{B}\}\).
For a steady star-formation rate \(\dot{M}_\star\), the corresponding instantaneous Galactic inventory of Earth-analogue HZ planets is obtained by dimensional conversion,
\begin{equation}
N_{{\rm HZ,MW},m}(>M_{\rm min})
= 10^{6}\,\eta_{\oplus,m}\,
\left(\frac{\dot{M}_\star}{M_\odot~{\rm yr}^{-1}}\right)\,
\overline{Y}_m(M_{\rm min}),
\label{eq:NMW_from_Ybar}
\end{equation}
where the factor \(10^{6}\) converts Myr to yr and \(\eta_{\oplus,m}\) is an effective occurrence factor for Earth analogues under multiplicity Method \(m\).
The cumulative contribution from stars above a threshold mass is then
\begin{equation}
N_{{\rm HZ,MW},m}(\ge M_{\rm cut}) = f_{\ge M_{\rm cut},m}\,N_{{\rm HZ,MW},m}(>M_{\rm min}),
\label{eq:NMW_from_f}
\end{equation}
with \(f_{\ge M_{\rm cut},m}\) defined in Equation~(\ref{eq:f_ge_Mcut_m}).
In this work we evaluate \(M_{\rm min}=0.8\,M_\odot\), and we treat \(\eta_{\oplus,m}\) as a scalar occurrence normalization. The absolute inventories reported below scale linearly with this parameter. For massive stars and wide-orbit rocky planets, \(\eta_{\oplus,m}\) is not measured, so the Milky-Way numbers should be interpreted as normalized estimates rather than occurrence-rate predictions.

\begin{deluxetable*}{lrrrrrrrrrr}
\tablecaption{Habitable-zone (HZ) summary for the rotating solar-metallicity \textsc{GENEC} grid ($Z=0.014$, $v/v_{\rm crit}=0.4$).
For each model we report $M_{\rm ini}$, $M_{\rm fin}$, the main-sequence lifetime $t_{\rm MS}$, the post--main-sequence duration $t_{\rm post}$, and the cumulative HZ durations $\Delta t_{\rm HZ,MS}$ and $\Delta t_{\rm HZ,post}$ defined by $r_{\rm in}(t)<r_{\rm out,clim}(t)$ with $r_{\rm in}=\max(r_{\rm in,clim},r_{\rm wind},r_{\rm XUV})$ and $r_{\rm out}=r_{\rm out,clim}$. The mean radii $\langle r_{\rm in}\rangle$ and $\langle r_{\rm out}\rangle$ are time-weighted over the timesteps that satisfy the HZ criterion, reported separately for MS and post-MS phases.}
\label{tab:1}
\tablehead{
Model & $M_{\rm ini}$ & $M_{\rm fin}$ & $t_{\rm MS}$ & $t_{\rm post}$ & $\Delta t_{\rm HZ,MS}$ & $\Delta t_{\rm HZ,post}$ & $\langle r_{\rm in}\rangle_{\rm MS}$ & $\langle r_{\rm out}\rangle_{\rm MS}$ & $\langle r_{\rm in}\rangle_{\rm post}$ & $\langle r_{\rm out}\rangle_{\rm post}$ \\
 & [$M_\odot$] & [$M_\odot$] & [Myr] & [Myr] & [Myr] & [Myr] & [AU] & [AU] & [AU] & [AU] \\
}
\startdata
P0p8Z14S0.4 & 0.800 & 0.796 & 22161.521 & 6786.888 & 21971.392 & 6786.888 & 0.594 & 1.020 & 1.017 & 1.746 \\
P0p9Z14S0.4 & 0.900 & 0.898 & 13627.331 & 4861.156 & 13579.799 & 4861.156 & 0.777 & 1.334 & 1.193 & 2.050 \\
P001Z14S0.4 & 1.000 & 0.998 & 8523.070 & 3835.352 & 8491.382 & 3835.352 & 0.982 & 1.686 & 1.418 & 2.436 \\
P002Z14S0.4 & 2.000 & 1.956 & 1284.293 & 94.348 & 1283.497 & 94.348 & 4.552 & 7.817 & 8.908 & 15.298 \\
P003Z14S0.4 & 3.000 & 2.983 & 403.946 & 48.270 & 403.683 & 48.270 & 10.258 & 17.616 & 11.835 & 20.325 \\
P004Z14S0.4 & 4.000 & 3.965 & 188.782 & 38.578 & 188.655 & 38.578 & 17.701 & 30.399 & 23.092 & 39.658 \\
P005Z14S0.4 & 5.000 & 4.940 & 108.861 & 19.970 & 108.779 & 19.970 & 26.663 & 45.790 & 37.050 & 63.629 \\
P007Z14S0.4 & 7.000 & 6.868 & 50.811 & 7.758 & 50.775 & 7.758 & 48.311 & 82.967 & 70.112 & 120.409 \\
P009Z14S0.4 & 9.000 & 8.517 & 31.082 & 4.131 & 31.058 & 4.131 & 74.021 & 127.123 & 112.872 & 193.843 \\
P012Z14S0.4 & 12.000 & 10.224 & 18.278 & 2.291 & 1.150 & 2.206 & 256.046 & 263.236 & 184.113 & 313.684 \\
P015Z14S0.4 & 14.999 & 11.071 & 13.372 & 1.583 & 0.000 & 1.485 & \nodata & \nodata & 269.231 & 461.641 \\
P020Z14S0.4 & 19.998 & 7.179 & 9.451 & 0.936 & 0.000 & 0.596 & \nodata & \nodata & 380.618 & 649.631 \\
P025Z14S0.4 & 24.995 & 9.690 & 7.858 & 0.686 & 0.000 & 0.215 & \nodata & \nodata & 499.339 & 855.779 \\
P032Z14S0.4 & 31.990 & 10.125 & 6.601 & 0.579 & 0.000 & 0.036 & \nodata & \nodata & 1016.901 & 1170.374 \\
P040Z14S0.4 & 39.981 & 12.332 & 5.662 & 0.475 & 0.000 & 0.000 & \nodata & \nodata & \nodata & \nodata \\
P060Z14S0.4 & 59.950 & 17.981 & 4.466 & 0.366 & 0.000 & 0.000 & \nodata & \nodata & \nodata & \nodata \\
P085Z14S0.4 & 84.901 & 26.393 & 3.715 & 0.327 & 0.000 & 0.000 & \nodata & \nodata & \nodata & \nodata \\
\enddata
\end{deluxetable*}


\section{Results}\label{sec:results}
We begin by mapping each rotating solar-metallicity \textsc{GENEC} track onto a time dependent habitable zone (HZ) band by combining classical climate boundaries with atmospheric retention constraints from XUV irradiation and wind erosion. Section~\ref{subsec:hz_windows} addresses the necessary condition for habitability by quantifying when an HZ annulus exists, defined by $r_{\rm in}(t)<r_{\rm out,clim}(t)$. Table~\ref{tab:1} summarizes the full grid by reporting the MS and post-MS durations satisfying this criterion and the corresponding time-weighted mean radii. In Section~\ref{subsec:hz_residence} we will then address the stronger requirement that a planet at fixed semimajor axis can remain inside the evolving annulus for a continuous interval that is long enough to matter.

\subsection{Existence of habitable-zone bands along rotating tracks}\label{subsec:hz_windows}
\begin{figure*}
\centering
\includegraphics[width=\textwidth]{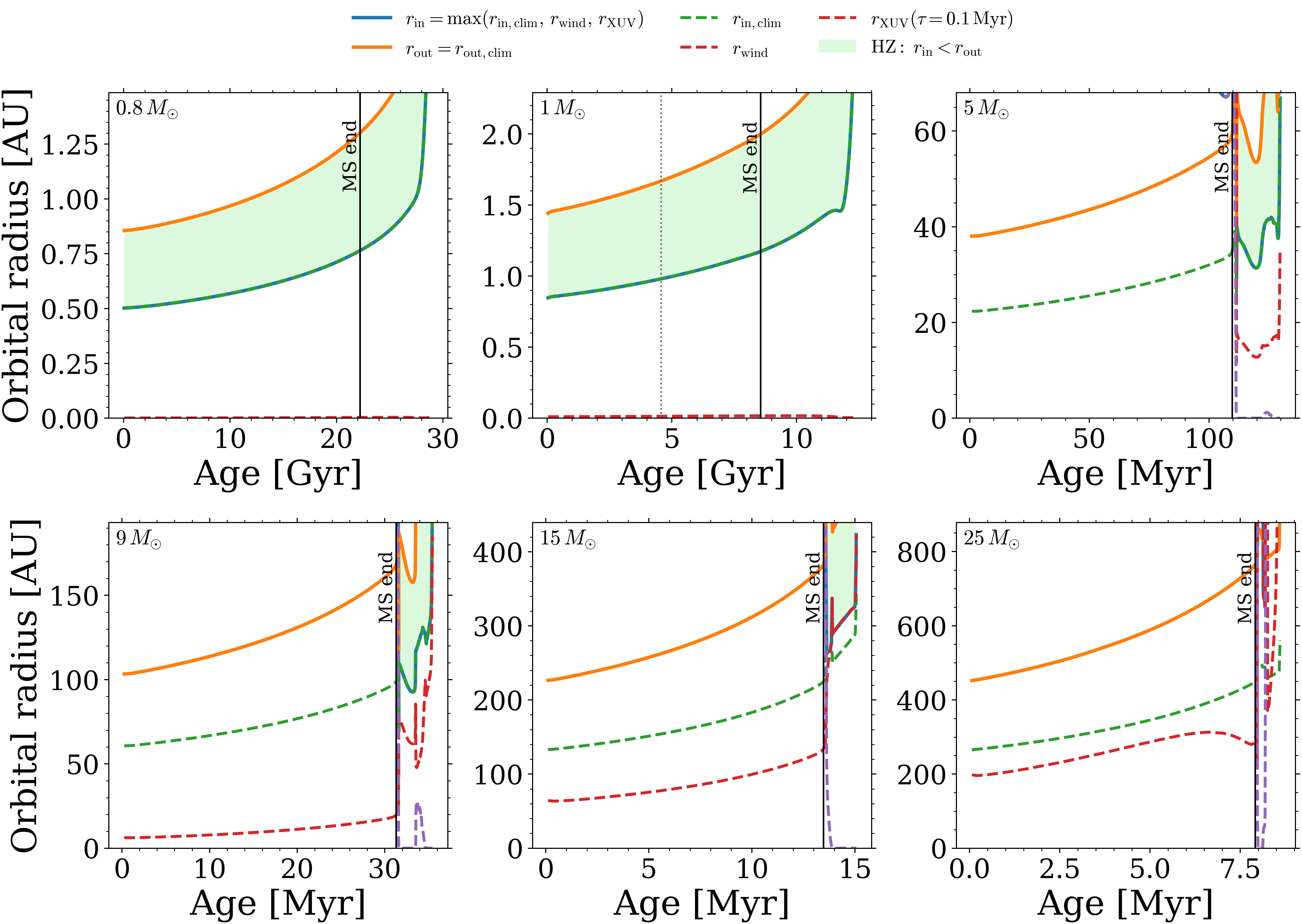}
\caption{
Time evolution of habitable-zone radii for six rotating solar-metallicity \textsc{GENEC} tracks ($0.8$, $1$, $5$, $9$, $15$, and $25\,M_\odot$; model identifiers are shown in the upper-left of each panel).
The climate-only inner and outer boundaries are $r_{\rm in,clim}$ (green dashed) and $r_{\rm out,clim}$ (orange), while the adopted inner edge is $r_{\rm in}=\max(r_{\rm in,clim},\,r_{\rm wind},\,r_{\rm XUV})$ (blue); $r_{\rm wind}$ (magenta dashed) and $r_{\rm XUV}$ (purple dashed, computed for $\tau=0.1\,\mathrm{Myr}$) denote the wind- and XUV-limited constraints on atmospheric retention.
The habitable band (light green shading) is defined by $r_{\rm in}<r_{\rm out,clim}$ at a given time.
Vertical black lines mark the end of core-H burning (MS end).
}
\label{fig:hz_band_sixpanel}
\end{figure*}

Figure~\ref{fig:hz_band_sixpanel} provides a guided view of the HZ evolution along representative tracks. Each panel shows the climate only edges $r_{\rm in,clim}(t)$ and $r_{\rm out,clim}(t)$ together with the adopted inner edge $r_{\rm in}(t)=\max(r_{\rm in,clim},r_{\rm wind},r_{\rm XUV})$. The shaded region marks epochs where $r_{\rm in}<r_{\rm out,clim}$. The vertical black line indicates the end of core-H burning. The low and intermediate mass cases are close to the climate boundary for much of the MS, while the transition to massive stars is controlled by the atmospheric retention terms as $r_{\rm in}$ approaches or exceeds $r_{\rm out,clim}$.

The $0.8$--$1\,M_\odot$ panels illustrate the low-mass regime where atmospheric loss constraints remain subdominant and the HZ is effectively set by the climate band. A continuous MS annulus is present for essentially the full core-H lifetime, with $\Delta t_{\rm HZ,MS}=21971\,\mathrm{Myr}$ for $t_{\rm MS}=22162\,\mathrm{Myr}$ at $0.8\,M_\odot$ and $\Delta t_{\rm HZ,MS}=8491\,\mathrm{Myr}$ for $t_{\rm MS}=8523\,\mathrm{Myr}$ at $1\,M_\odot$ (Table~\ref{tab:1}). The corresponding MS-averaged radii are $\langle r_{\rm in}\rangle_{\rm MS}=0.594$ and $\langle r_{\rm out}\rangle_{\rm MS}=1.020\,\mathrm{AU}$ at $0.8\,M_\odot$, and $\langle r_{\rm in}\rangle_{\rm MS}=0.982$ and $\langle r_{\rm out}\rangle_{\rm MS}=1.686\,\mathrm{AU}$ at $1\,M_\odot$. Post-MS habitability remains long-lived in this mass range, shifting outward to $\langle r_{\rm in}\rangle_{\rm post}=1.017\,\mathrm{AU}$ and $\langle r_{\rm out}\rangle_{\rm post}=1.746\,\mathrm{AU}$ over $\Delta t_{\rm HZ,post}=6787\,\mathrm{Myr}$ at $0.8\,M_\odot$ and to $\langle r_{\rm in}\rangle_{\rm post}=1.418\,\mathrm{AU}$ and $\langle r_{\rm out}\rangle_{\rm post}=2.436\,\mathrm{AU}$ over $\Delta t_{\rm HZ,post}=3835\,\mathrm{Myr}$ at $1\,M_\odot$.

The $5$--$9\,M_\odot$ panels shift the climate band to tens to hundreds of AU, reflecting the rapid rise in bolometric luminosity. An HZ annulus still exists for essentially the full MS, with $\Delta t_{\rm HZ,MS}=108.779\,\mathrm{Myr}$ for $t_{\rm MS}=108.861\,\mathrm{Myr}$ at $5\,M_\odot$ and $\Delta t_{\rm HZ,MS}=31.058\,\mathrm{Myr}$ for $t_{\rm MS}=31.082\,\mathrm{Myr}$ at $9\,M_\odot$ (Table~\ref{tab:1}). The characteristic MS radii increase from $\langle r_{\rm in}\rangle_{\rm MS}=26.663\,\mathrm{AU}$ and $\langle r_{\rm out}\rangle_{\rm MS}=45.790\,\mathrm{AU}$ at $5\,M_\odot$ to $\langle r_{\rm in}\rangle_{\rm MS}=74.021\,\mathrm{AU}$ and $\langle r_{\rm out}\rangle_{\rm MS}=127.123\,\mathrm{AU}$ at $9\,M_\odot$. Post-MS windows persist but shorten to $\Delta t_{\rm HZ,post}=19.970\,\mathrm{Myr}$ at $5\,M_\odot$ and $\Delta t_{\rm HZ,post}=4.131\,\mathrm{Myr}$ at $9\,M_\odot$, while the mean radii move outward to $\langle r_{\rm in}\rangle_{\rm post}=37.050\,\mathrm{AU}$, $\langle r_{\rm out}\rangle_{\rm post}=63.629\,\mathrm{AU}$ and to $\langle r_{\rm in}\rangle_{\rm post}=112.872\,\mathrm{AU}$, $\langle r_{\rm out}\rangle_{\rm post}=193.843\,\mathrm{AU}$, respectively. In this intermediate-mass regime (Figure~\ref{fig:hz_band_sixpanel}) the separation between constraints becomes visible, and the full-physics inner edge can depart from the climate inner edge as winds and high-energy irradiation begin to control atmospheric retention.

The bottom-middle and bottom-right panels ($15\,M_\odot$ and $25\,M_\odot$) demonstrate the high-mass outcome. Although a climate band exists at large radii, the MS HZ is absent for these models in Table~\ref{tab:1}, which implies $r_{\rm in}\ge r_{\rm out,clim}$ throughout core-H burning under the adopted XUV and wind scalings. Any remaining habitability is confined to brief post-MS intervals. At $15\,M_\odot$ the post-MS HZ persists for $\Delta t_{\rm HZ,post}=1.485\,\mathrm{Myr}$ with $\langle r_{\rm in}\rangle_{\rm post}=269.231\,\mathrm{AU}$ and $\langle r_{\rm out}\rangle_{\rm post}=461.641\,\mathrm{AU}$. At $25\,M_\odot$, the post-MS window shortens to $\Delta t_{\rm HZ,post}=0.215\,\mathrm{Myr}$ while shifting outward to $\langle r_{\rm in}\rangle_{\rm post}=499.339\,\mathrm{AU}$ and $\langle r_{\rm out}\rangle_{\rm post}=855.779\,\mathrm{AU}$. The persistence of a post-MS annulus at $15\,M_\odot$ but not beyond $\sim 1\,\mathrm{Myr}$ at higher mass motivates a practical threshold for existence based habitability.
Requiring at least $\sim 1\,\mathrm{Myr}$ of continuous annulus existence in any phase is satisfied at $15\,M_\odot$ but fails by $20\,M_\odot$ where $\Delta t_{\rm HZ,post}=0.596\,\mathrm{Myr}$, which places the transition near $\sim 18\,M_\odot$ by interpolation across the sampled grid. The table further shows that even post-MS habitability vanishes at higher masses, reaching $\Delta t_{\rm HZ,post}=0$ by $40\,M_\odot$ in this set. The limiting physics in the massive star regime is therefore the atmospheric retention inner boundary and its mass dependence, rather than the existence of a climate HZ at large orbital radii.

\subsection{Fixed-orbit residence times: feasibility beyond HZ existence}\label{subsec:hz_residence}

Section~\ref{subsec:hz_windows} established when an HZ annulus exists, meaning that at a given time there is a non-empty interval of radii satisfying $r_{\rm in}(t)<r<r_{\rm out,clim}(t)$.
A distinct question is feasibility for a single planet on a fixed orbit, because the HZ boundaries sweep outward as $L_\star(t)$ and the atmospheric-loss constraints evolve.
For a fixed semimajor axis $a$ we define the residence time as the longest contiguous time interval during which
$r_{\rm in}(t) < a < r_{\rm out,clim}(t)$ holds, and we then maximize over $a$.
We compute this maximum separately on the MS and post-MS by restricting the search to the corresponding time domains, and we denote the results by $\Delta t_{\rm res,MS}$ and $\Delta t_{\rm res,post}$.

\begin{figure}
\centering
\includegraphics[width=\columnwidth]{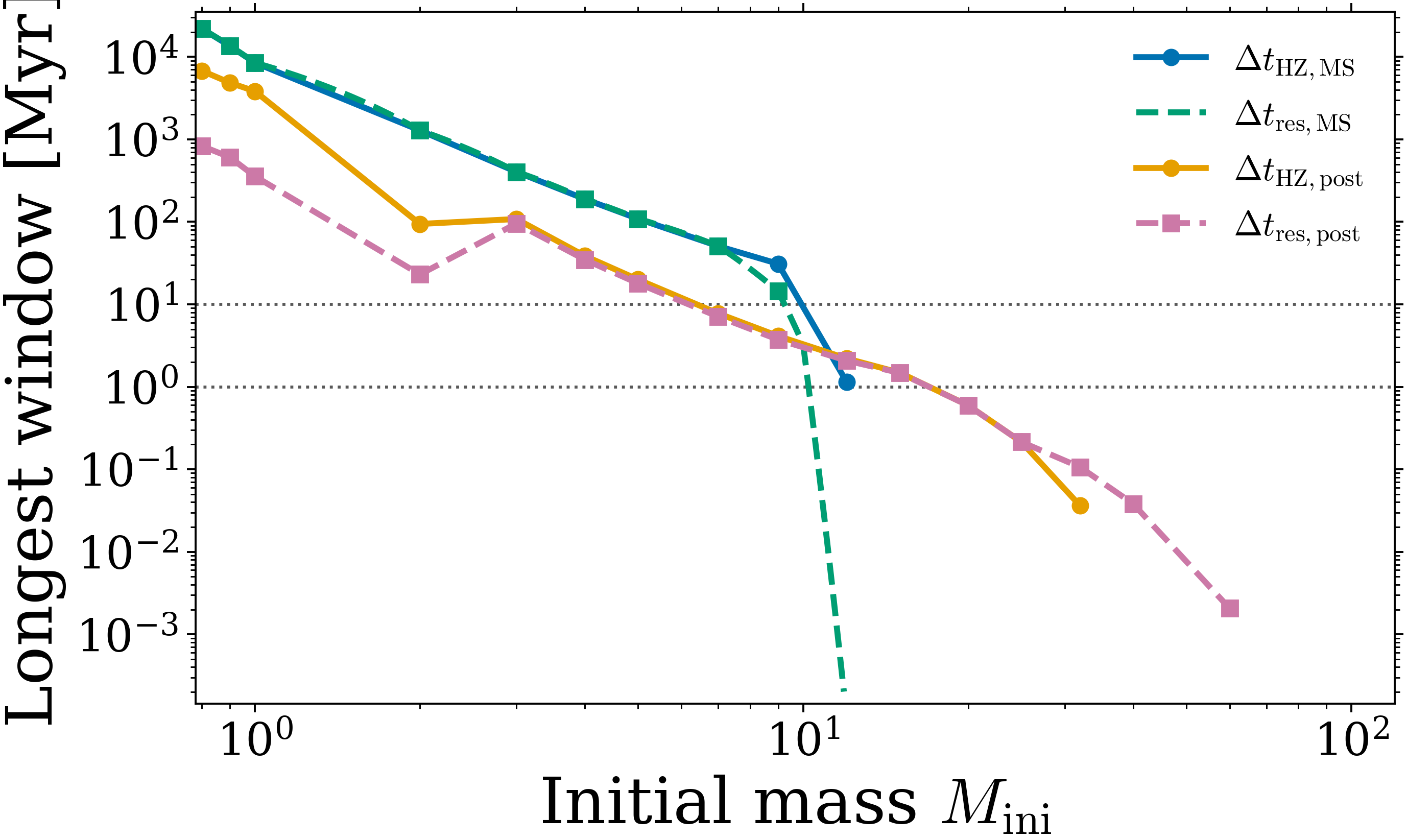}
\caption{
Existence time and maximum fixed-orbit residence time versus initial mass for the rotating solar-metallicity \textsc{GENEC} grid.
Solid curves show the cumulative HZ existence durations $\Delta t_{\rm HZ}$ and dashed curves show the maximum contiguous residence time $\Delta t_{\rm res}=\max_a \Delta t(a)$, each evaluated separately on the MS and post-MS.
Horizontal lines indicate benchmark residence requirements.}
\label{fig:existence_vs_residence}
\end{figure}

Figure~\ref{fig:existence_vs_residence} contrasts the existence times from Table~\ref{tab:1} with the corresponding residence maxima.
At low mass the two MS measures are nearly identical because the HZ band evolves slowly, so a broad range of $a$ remains inside continuously.
For example at $1\,M_\odot$ the MS existence time is $\Delta t_{\rm HZ,MS}=8.49\times 10^{3}\,{\rm Myr}$ and the optimal residence time tracks it closely in Fig.~\ref{fig:existence_vs_residence}.
At intermediate masses, the same remains true in an absolute sense even though the clock is faster.
At $5\,M_\odot$ we have $\Delta t_{\rm HZ,MS}=1.09\times 10^{2}\,{\rm Myr}$, and by $9\,M_\odot$ it is $\Delta t_{\rm HZ,MS}=3.11\times 10^{1}\,{\rm Myr}$, with $\Delta t_{\rm res,MS}$ remaining comparable because the MS band still sweeps outward smoothly enough that one can choose an orbit that stays inside for most of the MS.

The divergence between existence and residence emerges in the transition regime where atmospheric retention and rapid structural evolution compress the usable band and increase its sweep rate.
The relevant control parameter is not only the shrinking MS lifetime but also the growth and variability of $r_{\rm in}(t)=\max(r_{\rm in,clim},r_{\rm wind},r_{\rm XUV})$, which steepens with stellar mass as winds strengthen and the spectrum hardens.
As a result, $r_{\rm in}(t)$ can approach $r_{\rm out,clim}(t)$ and the band can become both narrow and fast-moving, so the inequality $r_{\rm in}(t) < a < r_{\rm out,clim}(t)$ cannot be maintained for long at any fixed $a$.
This is why the MS residence curve in Fig.~\ref{fig:existence_vs_residence} drops sharply beyond the point where the MS band is still present but no longer quasi-stationary.
A concrete example is the $12\,M_\odot$ track, which still has a non-zero MS existence time in Table~\ref{tab:1} ($\Delta t_{\rm HZ,MS}=1.15\,{\rm Myr}$), yet Fig.~\ref{fig:existence_vs_residence} shows that the maximum contiguous MS residence time is already pushed below the $\sim 1\,{\rm Myr}$ benchmark.
In this sense, the residence framing tightens the practical mass ceiling for continuous MS habitability, because it requires not only the existence of an annulus but also that the annulus does not sweep past any fixed orbit too rapidly.

The post-MS behaviour is similar in trend but is set by even faster luminosity and temperature evolution, so residence and existence are typically closer to each other and both are short.
At $15\,M_\odot$ Table~\ref{tab:1} gives $\Delta t_{\rm HZ,post}=1.49\,{\rm Myr}$ and Fig.~\ref{fig:existence_vs_residence} indicates an optimal post-MS residence time of the same order, whereas by $25\,M_\odot$ the post-MS existence time is $\Delta t_{\rm HZ,post}=2.15\times 10^{-1}\,{\rm Myr}$ and the corresponding residence maximum is comparably brief.
Thus, even when a post-MS HZ annulus exists, the rapid outward sweep of the boundaries limits any fixed-orbit residence to $\lesssim{\rm Myr}$ scales for massive stars, reinforcing that feasibility is controlled by atmospheric-retention constraints and evolutionary sweep rates rather than by the presence of a climate band at large radii. The following sensitivity tests isolate the $7$--$20\,M_\odot$ transition regime in tabular form.

\begin{figure*}
\centering
\includegraphics[width=\textwidth]{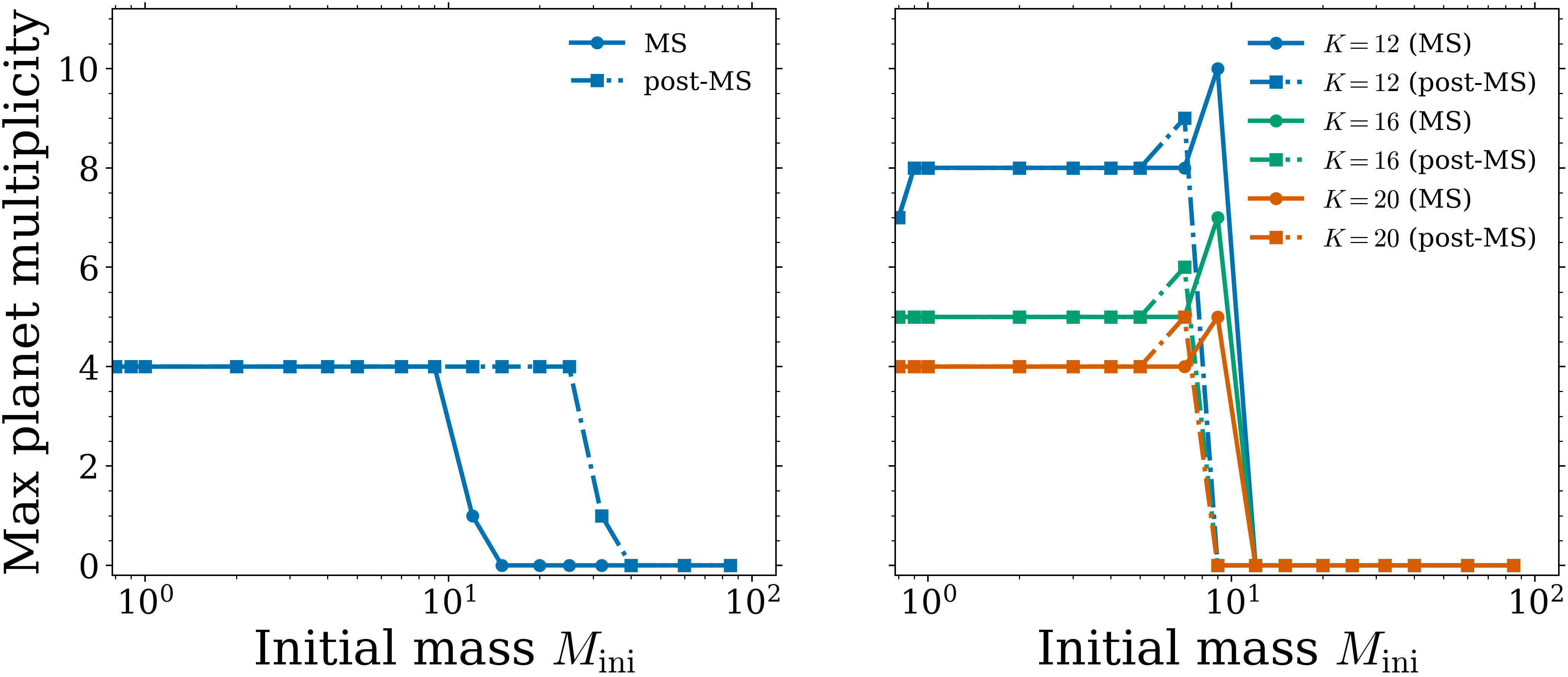}
\caption{Maximum HZ planet multiplicity inferred from Table~\ref{tab:1}. Solid lines denote MS values and dash--dotted lines denote post-MS values. Left: Method A (minimum period-ratio packing with $R=1.33$) gives $N_{\rm MS}=4$ for $0.8$--$9\,M_\odot$ and a sharp collapse at higher mass, while post-MS multiplicity persists to $25\,M_\odot$. Right: Method~B couples the HZ to a finite solids reservoir and shows $K=\{12,16,20\}$; it yields an MS plateau at low mass, an enhancement near $9\,M_\odot$, and rapid suppression once the HZ moves beyond the reservoir scale.}
\label{fig:hz_multiplicity}
\end{figure*}

\subsection{Sensitivity to the XUV normalization}
\label{subsec:xuv_sensitivity}

The residence-time ceiling depends on the normalization of the high-energy luminosity. We therefore repeated the calculation after multiplying the fiducial Planck-function XUV luminosity by a constant factor $f_{\rm XUV}$. This gives
\begin{equation}
L_{\rm XUV}\rightarrow f_{\rm XUV}L_{\rm XUV},
\qquad
r_{\rm XUV}\rightarrow f_{\rm XUV}^{1/2}r_{\rm XUV},
\label{eq:xuv_scaling}
\end{equation}
while keeping the stellar tracks and all other retention parameters fixed. This test isolates how the XUV normalization affects the transition-mass regime. Because Equation~\ref{eq:rXUV} depends on the product $L_{\rm XUV}\tau$, the same scaling can also be read as an exposure-time bracket at fixed XUV luminosity. For the fiducial $\tau=0.1$ Myr case, $f_{\rm XUV}=10$ and $100$ correspond to exposure-time requirements of $1$ and $10$ Myr, respectively. This does not evolve the atmospheric mass in time; it tests how the retention boundary moves when the required exposure time, or equivalently the integrated XUV dose, is varied.

Table~\ref{tab:xuv_scaling} shows the result for the common rotating and non-rotating mass grid $M_{\rm ini}=7,9,12,15,$ and $20\,M_\odot$. The fiducial case gives the same behavior described above. A main-sequence retention-limited annulus exists up to $12\,M_\odot$, but the longest fixed-orbit residence time above $1$ Myr is retained only up to $9\,M_\odot$. Thus the existence of an annulus remains a weaker condition than sustained residence at a fixed orbit.

The response to the XUV normalization is monotonic. Reducing $L_{\rm XUV}$ shifts the main-sequence ceiling upward. Increasing $L_{\rm XUV}$ shifts it downward. A factor of ten increase leaves the $1$ Myr residence ceiling at $9\,M_\odot$, while a factor of one hundred lowers it to $7\,M_\odot$. Reducing the XUV luminosity by factors of ten and one hundred moves the residence ceiling to $15$ and $20\,M_\odot$, respectively. The exact transition mass is therefore XUV-normalization dependent. The robust result is that atmospheric retention introduces a sharp main-sequence cutoff across the $7$--$20\,M_\odot$ transition regime. Stronger XUV forcing makes this cutoff more restrictive.

The post-main-sequence ceiling is less sensitive in this test. Across all five XUV scalings, a formal post-main-sequence annulus exists up to $20\,M_\odot$. The fixed-orbit residence ceiling above $1$ Myr remains at $15\,M_\odot$. In this mass range, the post-main-sequence result is therefore controlled mainly by the short evolutionary duration and rapid boundary motion rather than by the XUV normalization alone.

\begin{deluxetable*}{ccccc}
\tabletypesize{\scriptsize}
\tablecaption{
Sensitivity of the transition-mass ceiling to the XUV normalization for the common rotating and non-rotating mass grid $M_{\rm ini}=7,9,12,15,$ and $20\,M_\odot$. The listed values are the maximum sampled initial masses satisfying each criterion.
\label{tab:xuv_scaling}}
\tablehead{
\colhead{$f_{\rm XUV}$} &
\colhead{$M_{\rm max,MS}^{\rm exist}$} &
\colhead{$M_{\rm max,MS}^{\rm res\geq1\,Myr}$} &
\colhead{$M_{\rm max,post}^{\rm exist}$} &
\colhead{$M_{\rm max,post}^{\rm res\geq1\,Myr}$} \\
\colhead{} &
\colhead{[$M_\odot$]} &
\colhead{[$M_\odot$]} &
\colhead{[$M_\odot$]} &
\colhead{[$M_\odot$]}
}
\startdata
0.01 & 20 & 20 & 20 & 15 \\
0.10 & 15 & 15 & 20 & 15 \\
1.00 & 12 &  9 & 20 & 15 \\
10.0 &  9 &  9 & 20 & 15 \\
100  &  7 &  7 & 20 & 15 \\
\enddata
\end{deluxetable*}

\subsection{Sensitivity to the climate-boundary prescription}
\label{subsec:climate_sensitivity}

The fixed bolometric climate thresholds provide a simple fiducial mapping from stellar luminosity to HZ radius. This choice avoids extrapolating climate polynomials beyond their calibrated effective-temperature range. However, the climate boundary can still affect the marginal transition cases. We therefore repeated the transition-mass calculation using a Kopparapu-style temperature correction with the stellar effective temperature clipped at the hot edge of the calibration range. This gives a controlled estimate of the largest correction that can be applied without formally extrapolating the polynomial into the OB-star regime.

Table~\ref{tab:climate_sensitivity} shows the result for the same common rotating and non-rotating mass grid used in the XUV test. The clipped correction moves the main-sequence climate boundaries inward by about $11$--$12$ percent, with median ratios $r_{\rm in,clim}/r_{\rm in,clim}^{\rm fid}=0.884$ and $r_{\rm out,clim}/r_{\rm out,clim}^{\rm fid}=0.888$. This is enough to remove the marginal $12\,M_\odot$ main-sequence existence case. The main-sequence residence ceiling remains unchanged at $9\,M_\odot$. Thus the climate prescription affects whether a very narrow annulus formally exists at the transition mass, but it does not change the stronger fixed-orbit residence result.

The post-main-sequence ceilings are also unchanged in this test. A formal post-main-sequence annulus exists up to $20\,M_\odot$ on the common grid, while the residence ceiling above $1$ Myr remains at $15\,M_\odot$. This supports the interpretation from Section~\ref{subsec:hz_residence}: the limiting factor in the transition regime is sustained residence inside a rapidly evolving retention-limited annulus, not only the exact placement of the bolometric climate edges.

We also tested a formal extrapolation of the Kopparapu polynomial to the actual effective temperatures of the massive-star models. This diagnostic case produced nonphysical effective fluxes for about half of the main-sequence timesteps in the common transition-mass grid. We therefore do not adopt the extrapolated polynomial as a fiducial climate prescription. A self-consistent treatment of Earth-like climates under OB-star spectral energy distributions would require dedicated atmosphere calculations and is beyond the scope of the present survey.

\begin{table*}[t]
\centering
\caption{
Sensitivity of the transition-mass ceiling to the climate-boundary prescription for the common rotating and non-rotating mass grid $M_{\rm ini}=7,9,12,15,$ and $20\,M_\odot$. Values are maximum sampled initial masses.
}
\label{tab:climate_sensitivity}
\footnotesize
\begin{tabular}{lcccc}
\hline
Climate prescription &
$M_{\rm max,MS}^{\rm exist}$ &
$M_{\rm max,MS}^{\rm res>1}$ &
$M_{\rm max,post}^{\rm exist}$ &
$M_{\rm max,post}^{\rm res>1}$ \\
&
$[M_\odot]$ &
$[M_\odot]$ &
$[M_\odot]$ &
$[M_\odot]$ \\
\hline
Fixed bolometric thresholds & 12 & 9 & 20 & 15 \\
Kopparapu correction clipped at hot edge & 9 & 9 & 20 & 15 \\
\hline
\end{tabular}
\end{table*}
\subsection{Habitable-zone planet multiplicity}\label{sec:hz_multiplicity}

Figure~\ref{fig:hz_multiplicity} converts the phase-averaged HZ annuli in Table~\ref{tab:1} into a maximum number of planets that can be simultaneously accommodated within the HZ on the MS and post-MS. The calculation is a packing problem. For a given annulus, the multiplicity scales with the available logarithmic radial span, $\ln(\langle r_{\rm out}\rangle/\langle r_{\rm in}\rangle)$, divided by the logarithmic spacing imposed by the adopted mutual-Hill stability criterion (parameterized by $K$). Integer plateaus therefore arise naturally whenever the HZ log-width varies slowly with mass, even though the HZ radii themselves change by orders of magnitude.

\textit{Method A: geometric packing only.} In the simplest limit (left panel), we ignore disk truncation and solids budgets and treat the HZ annulus as continuously available for packing. This yields a nearly mass-invariant MS plateau with $N_{\rm MS}=4$ from $0.8$ to $9\,M_\odot$, followed by a rapid collapse to $N_{\rm MS}=1$ at $12\,M_\odot$ and $N_{\rm MS}=0$ for $M_{\rm ini}\ge 15\,M_\odot$. The post-MS curve remains at $N_{\rm post}=4$ through $25\,M_\odot$, then drops to $N_{\rm post}=1$ at $32\,M_\odot$ and vanishes for $M_{\rm ini}\ge 40\,M_\odot$. The panel is intentionally first-order in its physical interpretation. For low and intermediate masses, climate-based HZ edges scale approximately as $r\propto \sqrt{L_\star}$, so the logarithmic width remains of order unity. At higher masses, enhanced high-energy irradiation and winds shift the effective inner edge outward, reducing the annulus to the point that it is geometrically too narrow (or disappears) for multiple Hill-stable orbits.

\textit{Method B: disk-coupled packing with a finite solids reservoir.} The right panel adopts a more realistic architecture constraint by coupling the HZ to a finite solids reservoir with an outer disk scale and a radial solids weighting. This is motivated by standard disk models in which solids surface densities follow a declining power law and disks are truncated at $\mathcal{O}(10^2)\,\mathrm{AU}$ by formation conditions, viscous evolution, and photoevaporation. We show three stability spacings, $K=\{12,16,20\}$, where larger $K$ enforces wider separations and therefore lowers $N$ at fixed annulus width. The MS trend is a plateau, transition enhancement, and collapse sequence with explicit values: for $0.8$--$7\,M_\odot$ we obtain $N_{\rm MS}=(7,5,4)$ at $0.8\,M_\odot$ and $N_{\rm MS}=(8,5,4)$ from $0.9$ to $7\,M_\odot$ for $(K12,K16,K20)$, then an enhancement at $9\,M_\odot$ to $N_{\rm MS}=(10,7,5)$, and finally $N_{\rm MS}=0$ for $M_{\rm ini}\ge 12\,M_\odot$. The post-MS sequence is similar but terminates earlier: $N_{\rm post}=(7,5,4)$ at $0.8\,M_\odot$, $N_{\rm post}=(8,5,4)$ from $0.9$ to $5\,M_\odot$, a modest enhancement at $7\,M_\odot$ to $N_{\rm post}=(9,6,5)$, and $N_{\rm post}=0$ for $M_{\rm ini}\ge 9\,M_\odot$.

The resulting multiplicities are broadly consistent with previous dynamical HZ-packing studies. \citet{Obertas2017} found that tightly packed Earth-mass systems can remain stable for long integrations when adjacent planets are separated by several mutual Hill radii, and that five Earth-mass planets can fit in the Solar HZ under their adopted stability criterion. \citet{Kane2020} found that dynamically allowed HZ packing can reach roughly six Earth-mass planets for stellar masses above about $0.7\,M_\odot$, with the presence of giant planets strongly reducing the available stable real estate. Our Method A value of $N_{\rm MS}=4$ for $0.8$--$9\,M_\odot$ and our Method B values of order a few to several planets are therefore in the same range. The new element here is not the packing criterion itself, but its application to retention-limited HZs that evolve along massive-star tracks.

The enhancement near $9\,M_\odot$ in Method~B should be interpreted as a transition-grid feature rather than as a sharply resolved physical spike. It marks the last sampled MS model whose HZ still overlaps the adopted outer disk scale. Table~\ref{tab:1} shows that the $9\,M_\odot$ MS HZ spans $\langle r_{\rm in}\rangle_{\rm MS}\simeq 74\,\mathrm{AU}$ to $\langle r_{\rm out}\rangle_{\rm MS}\simeq 127\,\mathrm{AU}$, placing the annulus near the reservoir boundary. With finer mass sampling or a different $a_{\max}$, this feature would likely smooth into a transition. The physical point is therefore not the exact location or sharpness of the enhancement, but the rapid loss of disk overlap once the HZ moves beyond the assumed reservoir scale.

The combined message of Fig.~\ref{fig:hz_multiplicity} is therefore twofold. First, stellar evolution sets when an HZ exists and how wide it is, which explains the geometric plateaus and the disappearance of MS habitability above $\simeq 12\,M_\odot$. Second, once disk coupling is enforced, the dominant uncertainty shifts to the assumed availability of solids at wide separations. The multiplicity estimates should therefore be interpreted as dynamical capacity estimates for an assumed supply of terrestrial planets, not as predictions of planet formation. Beyond the transition regime, the limiting factor is no longer only the star's ability to host an HZ, but whether rocky planets can form or survive at the relevant orbital radii.


\subsection{Rotation effects on habitable zone evolution}
\label{subsec:rotation_hz}

\begin{figure*}
\centering
\includegraphics[width=\textwidth]{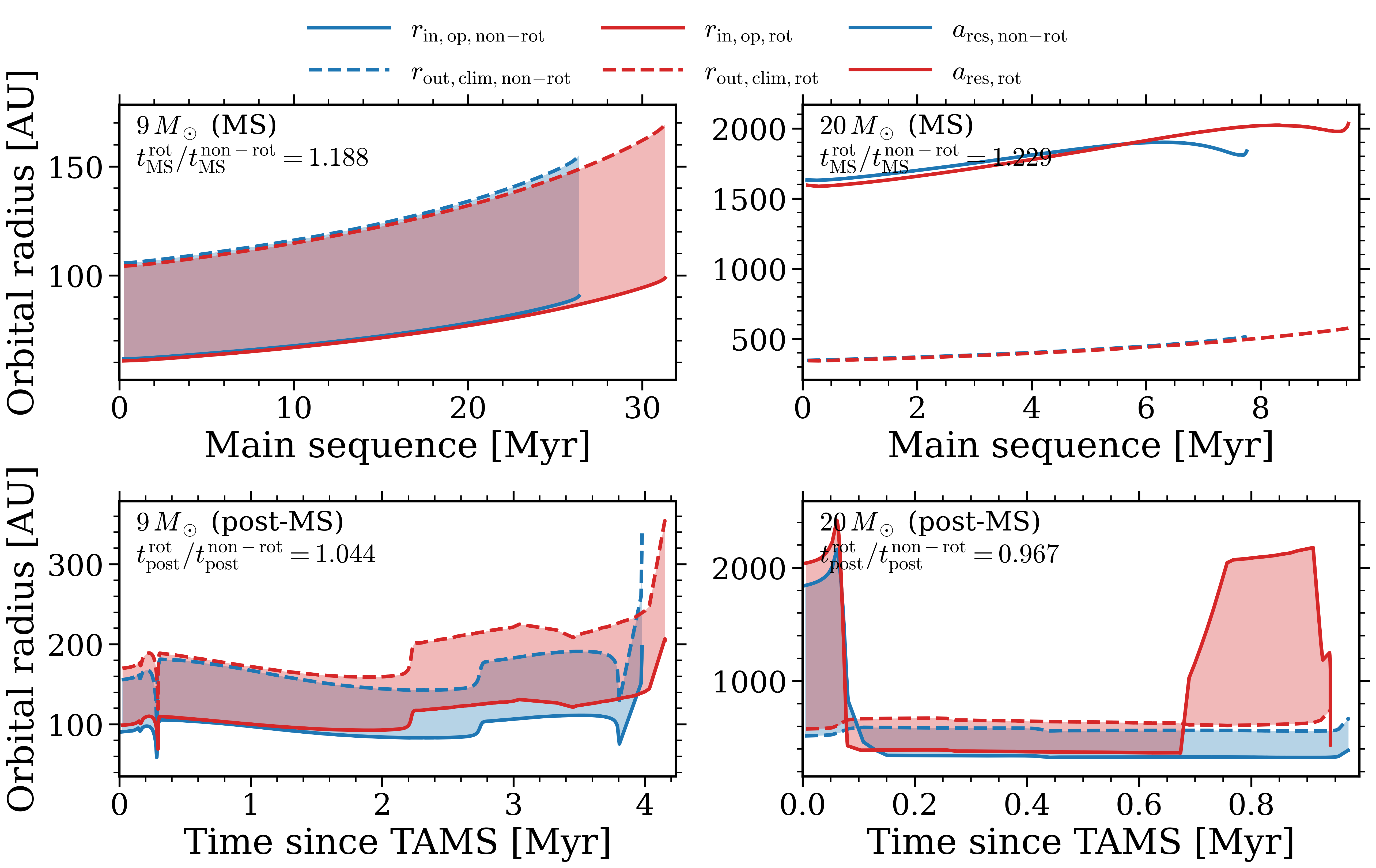}
\caption{
Rotation versus non-rotation comparison of HZ boundary evolution at $9\,M_\odot$ (left) and $20\,M_\odot$ (right).
Blue curves show non-rot \texttt{S0} models and red curves show rot \texttt{S0.4} models.
Solid curves plot the retention-limited inner edge $r_{\rm in}(t)$ and dashed curves plot the climate outer edge $r_{\rm out,clim}(t)$.
Thin horizontal lines mark the residence-orbit locations $a_{\rm res}$ in each phase.
Top panels use MS age.
Bottom panels use time since TAMS.
Shading marks epochs where an annulus exists, with overlap between rot and non-rot shown in purple.
The annotated ratios give $t_{\rm MS,rot}/t_{\rm MS,nonrot}$ (top) and $t_{\rm post,rot}/t_{\rm post,nonrot}$ (bottom).
}
\label{fig:D1_rotation}
\end{figure*}

We quantify rotation sensitivity by comparing models at fixed $M_{\rm ini}$ and $Z=0.014$ from the rotating grid (\texttt{S0.4}, ``rot'') and the non-rotating grid (\texttt{S0}, ``non-rot''). The HZ boundaries follow Section~\ref{subsec:hzmodel}. The outer edge is the climate boundary $r_{\rm out,clim}(t)$ from Equation~\ref{eq:climate}. The inner edge of the retention-limited HZ is $r_{\rm in}(t)$ from Equation~\ref{eq:rin}. Main-sequence (MS) and post-MS phases are separated using the $X_{c,{\rm H}}$ criterion in Section~\ref{subsec:tracks}. In Figure~\ref{fig:D1_rotation} the MS panels use stellar age, while the post-MS panels use time since TAMS. A habitable annulus exists when $r_{\rm in,op}(t)<r_{\rm out,clim}(t)$. When a phase contains no habitable timesteps we set $\Delta t_{\rm HZ}=0$ and we treat $a_{\rm res}$ as undefined for that phase.

At $9\,M_\odot$ on the MS (Figure~\ref{fig:D1_rotation}, top-left), both rot and non-rot models maintain a broad annulus that drifts outward with age. The inner edge rises from $\sim 65~{\rm AU}$ to $\sim 90~{\rm AU}$ over the non-rot MS, while the climate outer edge rises from $\sim 105~{\rm AU}$ to $\sim 150~{\rm AU}$. The rotating track follows nearly the same loci at fixed age, but it extends to later times and slightly larger terminal radii. Table~\ref{tab:1} gives $\langle r_{\rm in}\rangle_{\rm MS}=74.0~{\rm AU}$ and $\langle r_{\rm out}\rangle_{\rm MS}=127.1~{\rm AU}$ for the rotating model. The MS width is therefore typically $r_{\rm out,clim}-r_{\rm in}\sim 40$--$70~{\rm AU}$. Rotation mainly changes the clock. The annotated ratio gives $t_{\rm MS,rot}/t_{\rm MS,nonrot}=1.188$, so $t_{\rm MS,nonrot}\simeq 26.2~{\rm Myr}$ given $t_{\rm MS,rot}=31.08~{\rm Myr}$ in Table~\ref{tab:1}.

At $20\,M_\odot$ on the MS (Figure~\ref{fig:D1_rotation}, top-right), the climate band remains at large radii, with $r_{\rm out,clim}\sim 350$--$600~{\rm AU}$. The retention-limited inner edge is far larger, with $r_{\rm in}\sim 1.6$--$2.0\times 10^{3}~{\rm AU}$. Thus $r_{\rm in}$ exceeds $r_{\rm out,clim}$ by a factor of $\sim 3$--$5$ throughout core-H burning, so an MS annulus never forms for either model. The lifetime ratio is still substantial, with $t_{\rm MS,rot}/t_{\rm MS,nonrot}=1.229$ and $t_{\rm MS,nonrot}\simeq 7.69~{\rm Myr}$ from $t_{\rm MS,rot}=9.45~{\rm Myr}$ (Table~\ref{tab:1}). This has no impact on MS habitability because the limiting condition is the retention constraint rather than the climate band. At solar metallicity this is consistent with a regime where wind and irradiation terms become dominant at high mass, while rotation only perturbs the timing and the detailed trajectory in $L_\star(t)$ and $T_{\rm eff}(t)$.

Post-MS evolution at $9\,M_\odot$ (Figure~\ref{fig:D1_rotation}, bottom-left) preserves a sustained annulus across the full post-MS interval for both models. Table~\ref{tab:1} gives $\Delta t_{\rm HZ,post}=4.13~{\rm Myr}$ with $\langle r_{\rm in}\rangle_{\rm post}=112.9~{\rm AU}$ and $\langle r_{\rm out}\rangle_{\rm post}=193.8~{\rm AU}$ for the rotating track. The plotted boundaries fluctuate around these means, with widths typically of order $60$--$120~{\rm AU}$. Rotation changes the post-MS duration only mildly. The ratio $t_{\rm post,rot}/t_{\rm post,nonrot}=1.044$ implies $t_{\rm post,nonrot}\simeq 3.96~{\rm Myr}$ given $t_{\rm post,rot}=4.13~{\rm Myr}$. In this mass range the longer rotating clock maps directly into a modest increase in cumulative habitable time.

Post-MS evolution at $20\,M_\odot$ (Figure~\ref{fig:D1_rotation}, bottom-right) shows an early transient where $r_{\rm in}$ is initially very large and then collapses within $\sim 0.1~{\rm Myr}$ to values below $r_{\rm out,clim}$. During the mid post-MS interval the models sustain a wide annulus, with representative values $r_{\rm in,op}\sim 350$--$450~{\rm AU}$ and $r_{\rm out,clim}\sim 560$--$650~{\rm AU}$. Table~\ref{tab:1} gives $\langle r_{\rm in}\rangle_{\rm post}=380.6~{\rm AU}$ and $\langle r_{\rm out}\rangle_{\rm post}=649.6~{\rm AU}$ for the rotating model, consistent with the plotted mid-interval geometry. The key difference is late-time behavior. The rotating track exhibits a strong rise in $r_{\rm in}$ beginning near $(t-t_{\rm TAMS})\simeq 0.7~{\rm Myr}$, which pushes $r_{\rm in}$ well beyond $r_{\rm out,clim}$ and terminates the annulus. This indicates that the retention constraint becomes more restrictive late in the post-MS evolution. The figure alone does not identify which component of Equation(~\ref{eq:rin}) dominates the surge, but it is naturally associated with a rise in the wind- or XUV-limited contribution. The phase duration ratio is $t_{\rm post,rot}/t_{\rm post,nonrot}=0.967$, so rotation slightly shortens the post-MS clock at $20\,M_\odot$.

\begin{figure*}
\centering
\includegraphics[width=0.92\textwidth]{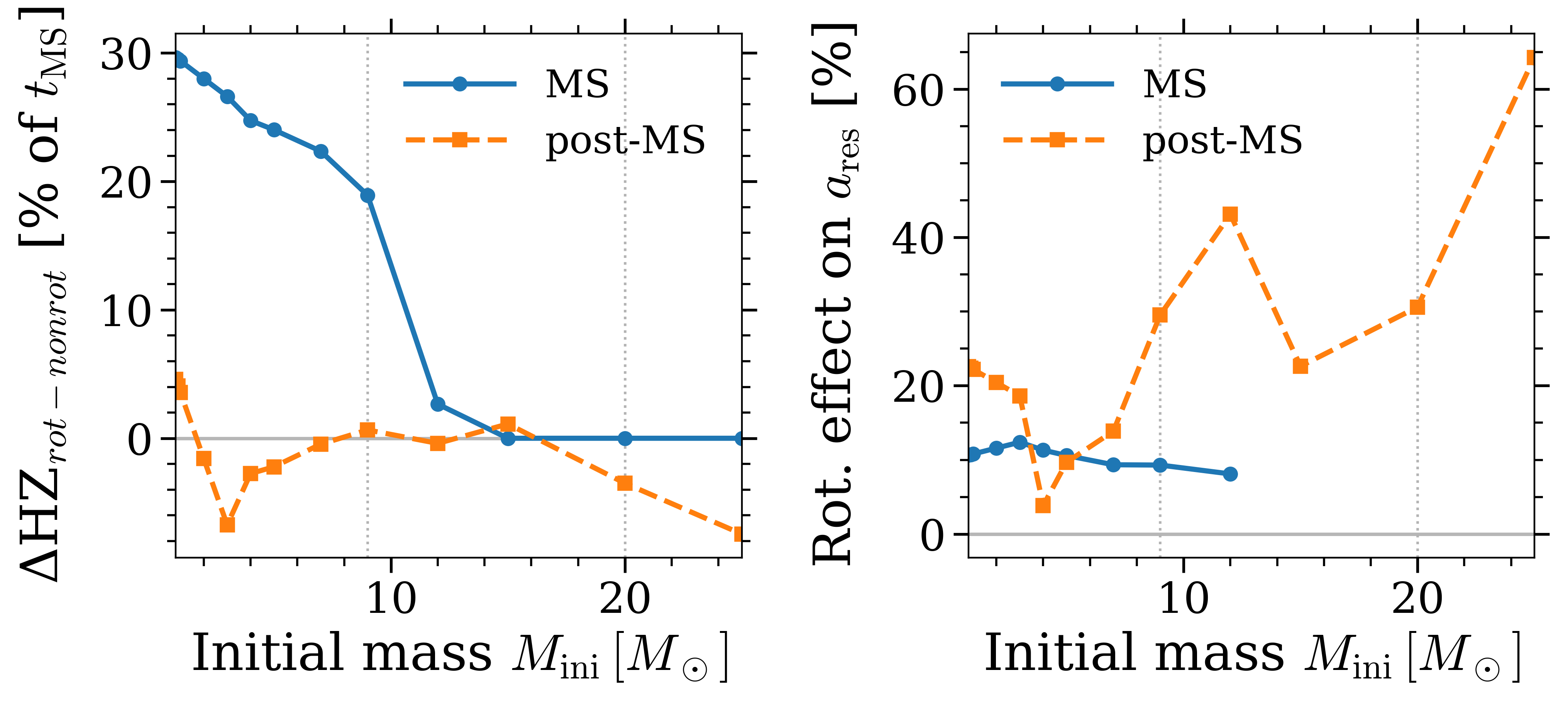}
\caption{Grid-wide rotation sensitivity of habitable time budgets and the residence orbit. Left panel shows $100\,(\Delta t_{\rm HZ,rot}-\Delta t_{\rm HZ,nonrot})/t_{\rm MS,nonrot}$ for MS and post-MS phases. Right panel shows $100\,[a_{\rm res,rot}/a_{\rm res,nonrot}-1]$ for MS and post-MS phases. Vertical dotted lines mark $M_{\rm ini}=9$ and $20\,M_\odot$, matching the detailed comparisons in Figure~\ref{fig:D1_rotation}.
}
\label{fig:D2_rotation}
\end{figure*}

Figure~\ref{fig:D2_rotation} summarizes the rotation response across the full grid. On the MS, rotation increases the cumulative habitable time at low and intermediate mass. The gain peaks near $9$--$10\,M_\odot$ at $\simeq 20\%$ of $t_{\rm MS,nonrot}$. This is consistent with Figure~\ref{fig:D1_rotation} because $\Delta t_{\rm HZ,MS}\approx t_{\rm MS}$ at $9\,M_\odot$ in Table~\ref{tab:1}. The MS signal collapses to zero by $\sim 15$--$20\,M_\odot$ because the MS annulus is absent in both grids. On the post-MS, the rotation-induced change in habitable time is modest at intermediate mass, and it becomes negative at $M\gtrsim 20\,M_\odot$. At $20$--$25\,M_\odot$ the reduction is a few percent of $t_{\rm MS,nonrot}$, which corresponds to a few $10^{-1}\,{\rm Myr}$, and it matches the earlier termination of the post-MS annulus for the rotating case in Figure~\ref{fig:D1_rotation}. The residence-orbit response is larger than the time budget response. On the MS, $a_{\rm res}$ shifts outward by $\sim 8$--$12\%$ where an MS HZ exists. On the post-MS, the shift reaches $\sim 30\%$ near $9$--$10\,M_\odot$ and rises to $\sim 60$--$65\%$ by $25\,M_\odot$. Across the solar-metallicity grid, rotation therefore acts mainly as a timing and orbit-selection effect at intermediate mass, while the existence of habitability at high mass is controlled by retention limits that are closely tied to mass loss.


\subsection{IMF-weighted contribution of massive stars}
The track-level analysis above shows that HZ annuli can exist at intermediate mass and can persist briefly at higher mass, but the population relevance depends on IMF weighting. We therefore fold the time-integrated yields $Y_{\rm A}(M_{\rm ini})$ and $Y_{\rm B}(M_{\rm ini};K,a_{\max})$ through several Milky-Way-like IMFs and evaluate two diagnostics on the initial-mass interval common to the non-rotating and rotating grids. We compute the IMF-integrated yield per unit stellar mass formed, $\overline{Y}$, and we quantify the massive star contribution with the fraction of the yield numerator contributed by $M_{\rm ini}\ge M_{\rm cut}$, denoted $f_{\ge M_{\rm cut}}$. We adopt $M_{\rm cut}=8\,M_\odot$ as the conventional massive star boundary. For Method~B we fix $K=16$ and bracket the uncertain outer disk scale with $a_{\max}=100$ and $1000\,{\rm AU}$.

\begin{figure}
\centering
\includegraphics[width=0.98\linewidth]{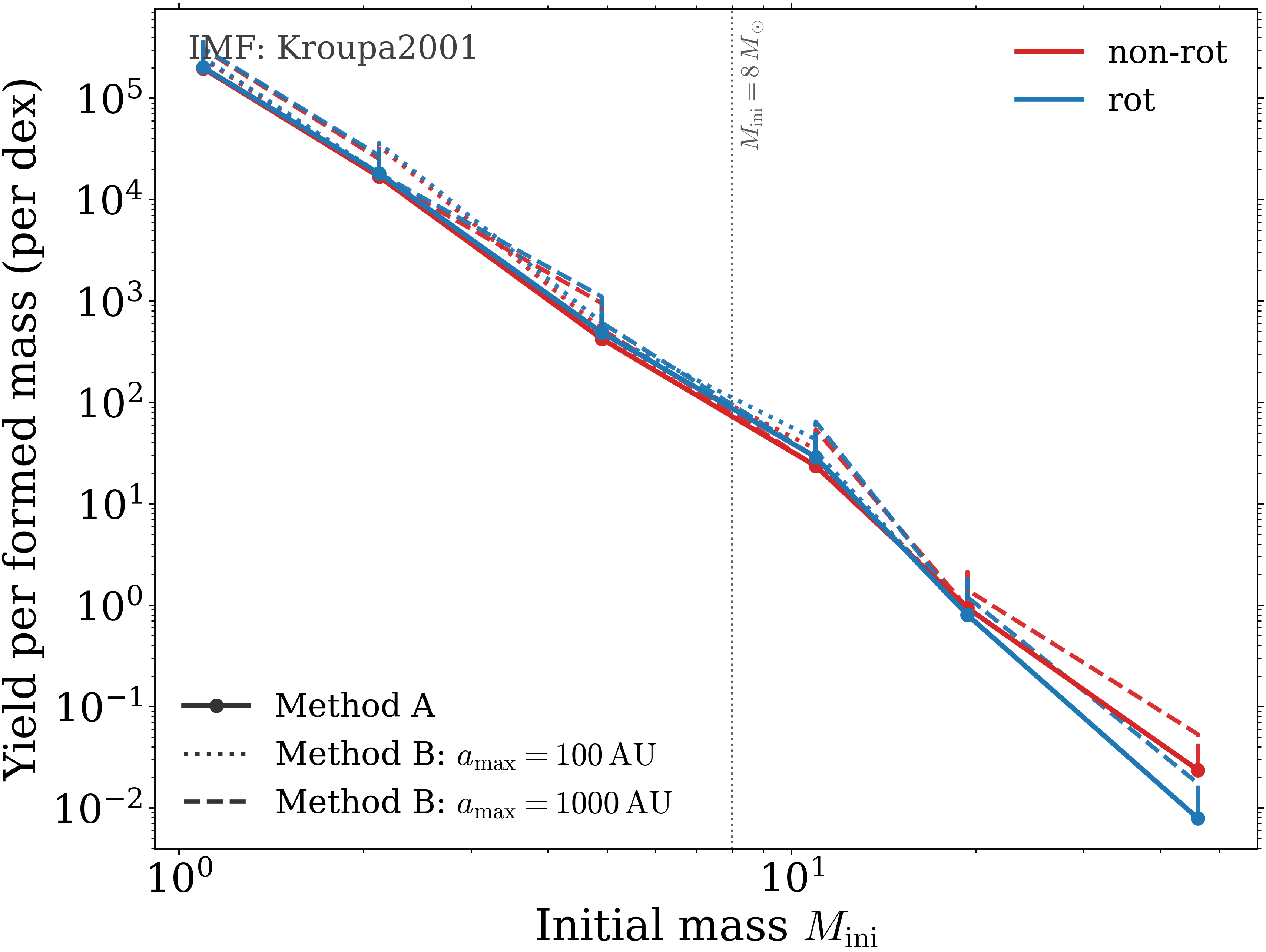}
\caption{
IMF-weighted habitability-yield density per logarithmic initial-mass interval (per dex), evaluated over MS+post-MS evolution on the initial-mass range common to the non-rotating (S0) and rotating (S0.4) grids.
Blue curves show S0.4 and red curves show S0.
Markers with solid lines show Method~A.
Dotted and dashed curves show Method~B for $a_{\max}=100\,{\rm AU}$ and $a_{\max}=1000\,{\rm AU}$, respectively, using the fiducial packing parameter $K=16$.
For clarity the plotted IMF is fixed to the Kroupa (2001) form; Salpeter and Chabrier IMFs preserve the qualitative mass ranking and are summarized in Table~\ref{tab:imf_metricA_option1}.
Points are plotted at the log-midpoint of each initial-mass bin, and the vertical dotted line marks $M_{\rm ini}=8\,M_\odot$.
}
\label{fig:imf_metricA_bins}
\end{figure}

Figure~\ref{fig:imf_metricA_bins} shows the IMF weighted habitability-yield density, $\mathrm{d}\overline{Y}/\mathrm{d}\log_{10}M_{\rm ini}$ (per dex), for the \citet{Kroupa2001} IMF, with color distinguishing S0 and S0.4 and line style distinguishing Method~A from Method~B. Each point gives the yield density associated with an initial-mass bin, plotted at the log-midpoint of the bin; multiplying by the bin width $\Delta\log_{10}M$ recovers that bin's contribution to the total $\overline{Y}$.
The mass ranking is strongly bottom-heavy because the lowest-mass tracks combine Gyr-scale HZ persistence with near-constant packing at low $M_{\rm ini}$, while higher-mass tracks evolve rapidly and their retention-limited HZ windows contract. Bins above a few solar masses lie at the $\lesssim 10^{-3}$ level in yield density relative to the low-mass bins on this scale.

Method~B tracks Method~A closely at low mass because the HZ typically remains well inside the adopted semimajor-axis domain. At higher mass the Method~B contribution is modestly reduced because the retention-limited boundaries sweep outward more rapidly and more often approach the imposed outer scale $a_{\max}$, which down-weights long-lived contributions at large radii. The difference between $a_{\max}=100$ and $1000\,{\rm AU}$ is therefore best interpreted as a controlled bracketing rather than a qualitative change in the mass ranking.

Table~\ref{tab:imf_metricA_option1} summarizes the corresponding IMF-integrated values of $\overline{Y}$ and the massive star numerator fraction $f_{\ge 8}$ for Salpeter, Kroupa, and Chabrier IMFs.
The rotating versus non-rotating difference in $\overline{Y}$ remains at the percent level on the common interval for both methods.
The massive star contribution is extremely small for all IMFs. For Method~B with $K=16$ we obtain $f_{\ge 8}\simeq(5$--$8)\times10^{-5}$, with only a few-percent change between $a_{\max}=100$ and $1000\,{\rm AU}$.
This robustness implies that even when HZ annuli occur at high mass, their contribution to the IMF-integrated planet--time budget is negligible under Milky-Way-like IMFs.

We now convert the IMF-integrated yields into an instantaneous Milky-Way inventory using the normalization in Section~\ref{sec:mw_norm}.
For a fiducial Milky-Way star-formation rate \(\dot{M}_\star=1.9\,M_\odot\,{\rm yr^{-1}}\) and an effective Earth-analogue occurrence factor \(\eta_\oplus=0.1\), we obtain \(N_{\rm HZ,MW}(M_{\rm ini}\ge0.8\,M_\odot)\simeq(2.4\text{--}2.8)\times10^{9}\) across the IMF choices in Table~\ref{tab:imf_metricA_option1}, with percent-level differences between S0 and S0.4. The value of \(\eta_\oplus\) is therefore a normalization choice. Changing it rescales the absolute inventory but does not change the IMF-weighted fraction \(f_{\geq 8}\) reported below.
This number is an order-of-magnitude estimate of the instantaneous Milky-Way inventory of Earth analogues that satisfy our adopted HZ and atmospheric-retention criteria.
It is not a time-integrated count over the Galaxy’s formation history.

The impact of extending the analysis to massive stars can be expressed using \(f_{\ge8}\).
Across Methods~A and~B we find \(f_{\ge8}\sim(0.6\text{--}1.3)\times10^{-4}\), so the massive star contribution is \(N_{\rm HZ,MW}(\ge8\,M_\odot)\simeq(1.5\text{--}3.5)\times10^{5}\) for the same fiducial \(\dot{M}_\star\) and \(\eta_\oplus\).

Thus including massive stars changes the Milky-Way total by only \(\Delta N/N \sim 10^{-4}\), because the IMF strongly favours low masses and because high-mass HZ windows are short even when they exist. In absolute terms, the extension adds of order a few \(10^{5}\) additional Earth analogues satisfying the adopted climate and retention filters at any instant under the fiducial occurrence normalization. This number is conditional on the formation or survival of rocky planets at the wide separations implied by massive-star HZs. If such planets are rare or absent around OB stars, then the effective occurrence factor for this population should be reduced accordingly, possibly to zero. Method~B reduces the high-mass tail relative to Method~A since enforcing \(a_{\max}=100\) or \(1000\,{\rm AU}\) suppresses multiplicity at high mass, but this difference only affects the already small \(\ge8\,M_\odot\) contribution and leaves the Galaxy-integrated total essentially unchanged.

\begin{deluxetable*}{llcccc}
\tablecaption{IMF-integrated habitability yield per unit stellar mass formed on the mass interval common to the rotating and non-rotating grids. We report the total yield $\overline{Y}$ for S0 and S0.4 for Method~A and for Method~B with $K=16$ at $a_{\max}=100\,{\rm AU}$ and $a_{\max}=1000\,{\rm AU}$. We also report the fraction of the yield numerator contributed by $M_{\rm ini}\ge 8\,M_\odot$, $f_{\ge 8}$, which isolates the massive star contribution independent of the IMF mass normalization.
}
\label{tab:imf_metricA_option1}
\tablehead{
\colhead{Method} & \colhead{IMF} & \colhead{$Y_{\rm non\text{-}rot}$} & \colhead{$Y_{\rm rot}$} & \colhead{$f_{\ge 8,\,\rm non\text{-}rot}$} & \colhead{$f_{\ge 8,\,\rm rot}$}
}
\startdata
Method A & Salpeter & $1.4\times10^{4}$ & $1.5\times10^{4}$ & $1.1\times10^{-4}$ & $1.3\times10^{-4}$ \\
Method A & Kroupa2001 & $1.3\times10^{4}$ & $1.4\times10^{4}$ & $1.1\times10^{-4}$ & $1.3\times10^{-4}$ \\
Method A & Chabrier2003 & $1.2\times10^{4}$ & $1.3\times10^{4}$ & $1.2\times10^{-4}$ & $1.4\times10^{-4}$ \\
Method B: $a_{\max}=100$ AU & Salpeter & $1.4\times10^{4}$ & $1.5\times10^{4}$ & $1.2\times10^{-4}$ & $1.5\times10^{-4}$ \\
Method B: $a_{\max}=100$ AU & Kroupa2001 & $1.3\times10^{4}$ & $1.4\times10^{4}$ & $1.2\times10^{-4}$ & $1.4\times10^{-4}$ \\
Method B: $a_{\max}=100$ AU & Chabrier2003 & $1.3\times10^{4}$ & $1.3\times10^{4}$ & $1.2\times10^{-4}$ & $1.5\times10^{-4}$ \\
Method B: $a_{\max}=1000$ AU & Salpeter & $1.4\times10^{4}$ & $1.5\times10^{4}$ & $1.3\times10^{-4}$ & $1.5\times10^{-4}$ \\
Method B: $a_{\max}=1000$ AU & Kroupa2001 & $1.3\times10^{4}$ & $1.4\times10^{4}$ & $1.3\times10^{-4}$ & $1.5\times10^{-4}$ \\
Method B: $a_{\max}=1000$ AU & Chabrier2003 & $1.3\times10^{4}$ & $1.3\times10^{4}$ & $1.3\times10^{-4}$ & $1.5\times10^{-4}$ \\
\enddata
\end{deluxetable*}

\section{Discussion}\label{sec:discussion}
\subsection{Dominant limiters and interpretation of IMF-weighted yields}
\label{subsec:dominant_limiters}

Figure~\ref{fig:imf_metricA_bins} shows that the IMF-weighted yield density is strongly concentrated at low initial mass, implying a rapidly declining cumulative contribution above any threshold $M_{\rm cut}$. The key quantitative outcome is $f_{\ge 8}\sim 10^{-4}$, with the exact value set by the IMF and weakly by rotation. Therefore massive stars contribute a negligible share of the total Earth-analogue planet--time budget under Milky-Way-like IMFs. This behaviour follows from IMF weighting combined with short high-mass windows and rapid boundary sweep, while the integral is dominated by long-lived low-mass tracks.

Method~A and Method~B differ only in the assumed availability of wide orbits. Method~A uses the full instantaneous annulus between $r_{\rm in}$ and $r_{\rm out,clim}$, then applies the packing prescription. Method~B enforces $a\le a_{\max}$ and tests how the small high-mass tail changes when very wide orbits are not freely available. These prescriptions do not model the formation of rocky planets at tens to hundreds of AU. They only quantify the dynamical capacity of the retention-limited annulus under assumed planet availability. The 100~AU and 1000~AU cases therefore mainly reshape the high-mass tail without changing the IMF-driven conclusion above.

The same scope applies to the wind-retention boundary. Equation~\ref{eq:rwind} uses a mean spherical wind and an assumed Earth-analogue magnetic shield. A weaker or absent dynamo would reduce magnetic protection, and sustained ram-pressure enhancements from wind structure would move $r_{\rm wind}$ outward according to Equation~\ref{eq:rwind_scaling}. These effects would make the retention-limited HZ more restrictive. They do not change the interpretation of the present calculation as a transparent screening model.

Rotation can shift track-level HZ timing and radii, but Table~\ref{tab:imf_metricA_option1} shows that $\overline{Y}$ changes at only the percent level between S0 and S0.4 because the IMF integral is controlled by low masses. Our integrals begin at $0.8\,M_\odot$, so the reported normalizations are conservative for a full Galactic inventory. When we translate $\overline{Y}$ into a Milky-Way number we are estimating an instantaneous quasi-steady inventory that scales with the present-day star-formation rate and the Earth-analogue occurrence factor, not a time-integrated count over Galactic history.

Before turning to detectability, it is useful to place our time scales in the context of previous work on luminous star and evolved star HZs. \citet{Ramirez2016} computed post MS HZs with one dimensional radiative convective climate models and stellar evolution tracks spanning $T_{\rm eff}=3700$ to $10{,}000\,{\rm K}$, corresponding roughly to M1 to A5 stars. They found that planets can remain in the post MS HZ for about $0.2$ to $9$ Gyr at solar metallicity. Our massive star post MS windows are much shorter. In the rotating grid, the post MS HZ duration is $4.13$ Myr at $9\,M_\odot$, $1.49$ Myr at $15\,M_\odot$, and $0.215$ Myr at $25\,M_\odot$. Even the $9\,M_\odot$ case is therefore about a factor of $50$ shorter than the lower end of the Ramirez and Kaltenegger post MS range, while the $15$ and $25\,M_\odot$ cases are shorter by factors of about $130$ and $900$. This difference follows from the much shorter evolutionary time scales of massive stars and from the rapid sweep of the retention constrained boundaries. \citet{Kaltenegger2021} provide a complementary observational context for HZ searches by considering which nearby stars can view Earth as a transiting planet. In that sense, prior work supports the basic expectation that luminous hosts move temperate orbits to larger separations. Additionally once XUV escape, wind pressure, and fixed orbit residence are imposed on massive star tracks. Therefore the existence of a bolometric climate HZ is not enough to produce long residence times.


\subsection{Detectability with the transit method: the long-period barrier}
\label{subsec:transit_detectability}

Our retention-limited HZ calculation places temperate orbits around intermediate and high mass stars at tens to hundreds of AU (Table~\ref{tab:1}). The immediate implication is simple: the corresponding orbital periods are far longer than the time baselines of transit surveys. Kepler's third law, written in a convenient scaling form, is
\begin{equation}
P \simeq 1~{\rm yr}\,
\left(\frac{a}{\rm AU}\right)^{3/2}
\left(\frac{M_\star}{M_\odot}\right)^{-1/2},
\label{eq:P_scaling_discussion}
\end{equation}
where $a$ is the semimajor axis. For $5\,M_\odot$, Table~\ref{tab:1} gives $\langle r_{\rm in}\rangle_{\rm MS}=26.7$~AU and $\langle r_{\rm out}\rangle_{\rm MS}=45.8$~AU. Taking a characteristic separation $a_{\rm mid}\equiv(\langle r_{\rm in}\rangle\langle r_{\rm out}\rangle)^{1/2}\approx 35$~AU yields $P\simeq 92$~yr. The effect is even more extreme at higher mass. In our fiducial $15\,M_\odot$ model the MS annulus vanishes, and the remaining post-MS window lies at $\langle r_{\rm in}\rangle_{\rm post}=269$~AU to $\langle r_{\rm out}\rangle_{\rm post}=462$~AU, implying $a_{\rm mid}\approx 3.5\times 10^2$~AU and $P\simeq 1.7\times 10^3$~yr.

These periods are simply incompatible with the cadence and duration of transit missions. \textit{TESS} typically monitors a field for $\sim$27~days (with longer coverage only near the continuous viewing zones), and \textit{Kepler}- and \textit{PLATO}-class surveys extend this to only a few years \citep[e.g.,][]{Ricker2015,Borucki2010,Rauer2014}. Multi-transit confirmation is therefore out of reach. Even a single-transit detection is strongly suppressed when $P\gg T_{\rm obs}$, because one must both align the orbit and ``catch'' the transit in the observing window:
\begin{equation}
p_{\rm 1tr}\sim \left(\frac{R_\star}{a}\right)\left(\frac{T_{\rm obs}}{P}\right).
\end{equation}
For the $5\,M_\odot$ case, adopting an illustrative MS radius $R_\star\sim 3R_\odot$ and a generous $T_{\rm obs}=4$~yr gives $p_{\rm 1tr}\sim 2\times 10^{-5}$. A single \textit{TESS} sector reduces this further by $\sim 27~{\rm d}/4~{\rm yr}\approx 1/54$. Finally, the signal itself becomes smaller as stars get larger: for terrestrial planets,
$\delta \approx (R_p/R_\star)^2 \approx 84~{\rm ppm}\,(R_\star/R_\odot)^{-2}$, placing Earth-analogue depths at a few ppm on the MS and well below 1~ppm for post-MS radii. In practice, variability in hot, massive stars only strengthens this conclusion. Taken together, these scalings imply that transits are not a practical discovery channel for temperate terrestrial planets in the massive star HZs predicted by our model.

\subsection{Direct detection at tens--$10^3$~AU: separation is favorable, contrast is not}
\label{subsec:direct_imaging}

Wide HZ orbits are a mixed case for direct detection. On the sky, they are easy to separate: $\theta\simeq a/d$ gives $\theta\simeq 0\farcs35$ for $a\simeq 35$~AU at $d=100$~pc, and $\theta\simeq 3\farcs5$ for $a\simeq 350$~AU. The challenge is not resolving the planet, but collecting enough photons against the glare of the host.

In reflected light, the planet/star flux ratio at phase angle $\alpha$ can be summarized by
\begin{equation}
C_{\rm ref}\ \sim\ A_g\,\Phi(\alpha)\,\left(\frac{R_p}{a}\right)^2,
\label{eq:reflected_contrast}
\end{equation}
so moving the HZ outward rapidly drives the contrast down as $a^{-2}$. For an Earth analogue with $A_g\Phi\simeq 0.1$, we find $C_{\rm ref}\sim 1.5\times 10^{-13}$ at $a\simeq 35$~AU and $C_{\rm ref}\sim 1.5\times 10^{-15}$ at $a\simeq 350$~AU. These levels sit orders of magnitude below the $\sim 10^{-10}$ contrast scale typically associated with reflected-light exoEarth imaging requirements of coronagraphs\citep[e.g.,][]{McElwain2025}. This steep penalty is the direct-imaging counterpart of the basic climate scaling: because the HZ expands roughly as $r\propto\sqrt{L_\star}$, it quickly moves into a regime where reflected-light detection of temperate terrestrial planets becomes prohibitive. However, one can search for HZ around massive stars with a few solar masses. 

Thermal emission in the mid-infrared avoids the explicit $a^{-2}$ suppression, but it trades contrast for sensitivity. A $T\sim 300$~K, $R_\oplus$ planet is intrinsically faint, and massive stars are rare enough that promising targets tend to be farther away. This is precisely the motivation for MIR interferometry concepts \citep[e.g.,][]{Quanz2022,dlr2022}. Moreover, for $M_\star\gtrsim 15\,M_\odot$ the habitable window in our grid is post-MS (Table~\ref{tab:1}). Any thermal-IR assessment should therefore use track-derived $R_\star(t_\star)$ and $T_{\rm eff}(t_\star)$ evaluated at the representative epoch defined in section~\ref{subsec:timediag}.

\subsection{How detectability scales with stellar mass: Sun-like hosts versus 5, 10, and 15~$M_\odot$}
\label{subsec:mass_scaling_detectability}

Around Sun-like stars, the classic picture holds: AU-scale HZ orbits yield $P\sim 1$~yr, transit probabilities of order $p_{\rm tr}\sim R_\star/a\sim 5\times 10^{-3}$, and Earth-analogue depths of $\delta\sim 84$~ppm. In reflected light, these systems define the familiar exoEarth benchmark $C_{\rm ref}\sim 10^{-10}$. In our rotating grid, increasing stellar mass primarily pushes the HZ outward (Table~\ref{tab:1}). That single shift drives the two key detection penalties at once: periods grow as $P\propto a^{3/2}$, while reflected-light contrast falls as $C_{\rm ref}\propto a^{-2}$.

For $5\,M_\odot$, the MS HZ spans $\sim 27$--$46$~AU, so $P\simeq 92$~yr and $C_{\rm ref}\sim 10^{-13}$ at $a\sim 35$~AU. The $\sim 10\,M_\odot$ regime lies near a sharp transition in our grid. At $9\,M_\odot$ the MS HZ spans $\sim 74$--$127$~AU, implying $P\simeq 3\times 10^2$~yr and $C_{\rm ref}\sim{\rm few}\times 10^{-14}$ at $a\sim 10^2$~AU. By $12\,M_\odot$ the MS annulus is both brief ($\Delta t_{\rm HZ,MS}=1.15$~Myr) and narrow ($\langle r_{\rm in}\rangle_{\rm MS}\simeq 256$~AU, $\langle r_{\rm out}\rangle_{\rm MS}\simeq 263$~AU), pushing characteristic periods to $P\sim 10^3$~yr. At $15\,M_\odot$ the MS HZ disappears entirely. Habitability is then confined to a short post-MS episode ($\Delta t_{\rm HZ,post}=1.49$~Myr) at $\sim 269$--$462$~AU, where $P\simeq 1.7\times 10^3$~yr and $C_{\rm ref}\sim 10^{-15}$.

The message is therefore direct. Once the HZ moves to tens to hundreds of AU, transits lose their leverage and reflected-light imaging becomes contrast limited. In our grid, the highest-mass cases also shift habitability to brief post-MS windows, which further motivates long wavelength direct detection approaches if such planets exist in nature.

\section{Summary and Conclusion}\label{sec:conclusion}

Massive stars dominate the radiative output of young stellar populations, but their winds and high-energy emission also push temperate climates to wide orbits where atmospheric retention is difficult. This paper quantifies where these two tendencies balance for Earth analogues at solar metallicity.
We coupled \textsc{GENEC} evolutionary tracks to climate boundaries and to atmospheric-retention limits, and we defined a retention-limited habitable annulus using $r_{\rm out}=r_{\rm out,clim}$ and $r_{\rm in}=\max(r_{\rm in,clim},r_{\rm wind},r_{\rm XUV})$.

We then characterized habitability in three complementary ways, the total time an annulus exists, the longest continuous residence time at any fixed orbit, and the maximum number of dynamically packed Earth analogues that can fit inside the annulus under two plausible orbit-availability prescriptions.
Finally, we folded these results through Milky-Way-like IMFs to estimate the instantaneous Galactic inventory of Earth analogues that satisfy the adopted climate and retention filters. The principal results are as follows:

\begin{enumerate}
\item
\emph{A sharp main-sequence ceiling appears near $\sim 10\,M_\odot$.}
At $9\,M_\odot$ a retention-limited MS annulus persists for $\Delta t_{\rm HZ,MS}=31.06$ Myr with characteristic radii $\sim 74$--$127$ AU, while by $12\,M_\odot$ it becomes a brief and extremely narrow episode with $\Delta t_{\rm HZ,MS}=1.15$ Myr and $\langle r_{\rm in}\rangle_{\rm MS}\simeq256$ AU and $\langle r_{\rm out}\rangle_{\rm MS}\simeq263$ AU.
By $15\,M_\odot$ no retention-limited MS annulus remains in our grid.

\item
\emph{Post-MS habitability survives to higher masses, but it is intrinsically short and wide-orbit.}
At $15\,M_\odot$ the post-MS window lasts $\Delta t_{\rm HZ,post}=1.49$ Myr and lies at $\sim 269$--$462$ AU.
At $25\,M_\odot$ it contracts to $\Delta t_{\rm HZ,post}=0.215$ Myr at $\sim 499$--$856$ AU.
A residual window persists at $32\,M_\odot$ for only $\Delta t_{\rm HZ,post}=0.036$ Myr at $\sim 10^{3}$ AU scales, and it vanishes by $40\,M_\odot$ in our models.

\item
\emph{Residence time is the limiting factor.}
In the transition regime, the HZ can exist yet migrate outward too quickly for sustained habitability at any fixed orbit. The $12\,M_\odot$ track demonstrates this: a finite MS existence time, but a longest contiguous MS residence time below $\sim 1$~Myr.

\item
\emph{Multiplicity is robust at low mass and collapses at high mass, and uncertainties in wide-orbit availability mainly reshape a negligible tail.}
The maximum packed multiplicity remains near-constant through $\sim 9\,M_\odot$ but it falls sharply once the retention-limited band becomes narrow and fast-moving, and it vanishes on the MS by $\ge 15\,M_\odot$.
Alternative assumptions about the availability of very wide orbits change the high-mass contribution modestly, but they do not alter the mass ranking because the relevant windows are short and rare.

\item
\emph{The Milky-Way inventory is set by low-mass hosts, while massive stars contribute a negligible fraction but a large absolute number.}
For $\dot{M}_\star=1.9\,M_\odot\,{\rm yr^{-1}}$ and an effective Earth-analogue occurrence factor $\eta_\oplus=0.1$, we obtain an instantaneous inventory of $N_{\rm HZ,MW}(M_{\rm ini}\ge 0.8\,M_\odot)\simeq(2.4$--$2.8)\times10^{9}$ Earth analogues satisfying our criteria. 
Across our bracketing assumptions we find \(f_{\ge 8}\sim(0.6\)--\(1.3)\times10^{-4}\). For the fiducial normalization \(\eta_\oplus=0.1\), this corresponds to \(N_{\rm HZ,MW}(\ge 8\,M_\odot)\simeq(1.5\)--\(3.5)\times10^{5}\). The absolute number scales linearly with \(\eta_\oplus\), which remains unconstrained for wide-orbit rocky planets around massive stars, while the fractional contribution remains only \(\Delta N/N\sim10^{-4}\).
\end{enumerate}

Across all of these metrics, the controlling failure mode at high mass is the loss of scale separation between the inner and outer edges, because $r_{\rm in}(t)$ steepens and becomes more time-variable with stellar mass until the inequality $r_{\rm in}(t)<r_{\rm out,clim}(t)$ is satisfied only intermittently and for rapidly sweeping, geometrically thin bands.

Detectability reflects the same geometry rather than the same demographics.
The large separations help angular resolution, but they impose extreme period and reflected-light penalties, with $C_{\rm ref}\sim 1.5\times10^{-13}$ at $a\simeq35$ AU and $C_{\rm ref}\sim 1.5\times10^{-15}$ at $a\simeq350$ AU for an Earth analogue.

In the follow-up work, we plan to venture from a Milky-Way snapshot to a cosmic history and to test whether the same retention-limited mass ceiling persists when stellar populations change.
Applying the framework to metal-poor tracks with evolving IMFs and cosmic star-formation histories will let us ask a sharp question, when do long-lived retention-limited annuli first become common enough to matter, and does the answer shift in environments where massive stars are more numerous.
This extension should be paired with planet models that depart from Earth analogues, because super-Earth gravities, higher atmospheric masses, and different volatile inventories can move the retention-limited inner edge and may broaden the narrow transition regime around $\sim 10$--$12\,M_\odot$.
The main physical uncertainty to tighten is the time dependence of the high-energy and wind forcing. A valuable test is to replace the adopted XUV and wind prescriptions with empirically calibrated, track-dependent histories and to propagate them through escape models that include atmospheric chemistry and magnetic protection.
Finally, the wide-orbit regime that dominates massive star HZ radii should be confronted with formation and survival physics in OB environments, including disk truncation, external photoevaporation, dynamical heating in clusters, and binarity. Binary evolution can alter the luminosity history, wind forcing, irradiation geometry, and dynamical stability of wide-orbit planets, but assessing these effects requires a separate binary stellar evolution and orbital dynamics treatment.

\begin{acknowledgments}
DN was supported by the Swiss National Science Fund (SNSF) Postdoctoral Fellowship, grant number: P500-2235464. AL was supported in part by the Black Hole Initiative at Harvard University, funded by grants from JTF and GBMF, and by the Galileo Project. We would like to thank the referee for their constructive comments and feedback during the review process. 
\end{acknowledgments}

\bibliography{sample7}{}
\bibliographystyle{aasjournalv7}



\end{document}